\RequirePackage{fix-cm}

\documentclass[twocolumn,natbib]{svjour3}          
\smartqed  
\usepackage{graphicx}
\usepackage{mathptmx}   

\usepackage{amsmath}
\usepackage{amssymb}
\usepackage{subfigure}
\usepackage{xfrac}
\usepackage{nicefrac}
\usepackage{tabu}
\usepackage{lscape}

\newcommand{\sref}[1]{Sect.~\ref{#1}}

\newcommand{\fref}[1]{Fig.~\ref{#1}}
\newcommand{\tref}[1]{Table~\ref{#1}}

\newcommand{\der}[3][]{\frac{d^{#1}#2}{d#3^{#1}}}
\newcommand{\Ion}[2]{\ensuremath{\left[\mathrm{#1}^{#2}\right]}}
\newcommand{\sciE}[1]{\ensuremath{\times10^{#1}}}
\newcommand{\INaF}{\ensuremath{I_{\mathrm{Na,F}}}}
\newcommand{\INaK}{\ensuremath{I_{\mathrm{Na,K}}}}
\newcommand{\INaB}{\ensuremath{I_{\mathrm{Na,B}}}}
\newcommand{\ICaB}{\ensuremath{I_{\mathrm{Ca,B}}}}
\newcommand{\IKA}{\ensuremath{I_{\mathrm{K,A}}}}
\newcommand{\IKD}{\ensuremath{I_{\mathrm{K,D}}}}
\newcommand{\IKdr}{\ensuremath{I_{\mathrm{K,dr}}}}
\newcommand{\IM}{\ensuremath{I_{\mathrm{m}}}}
\newcommand{\ICaP}{\ensuremath{I_{\mathrm{Ca,P}}}}
\newcommand{\INaCa}{\ensuremath{I_{\mathrm{Na,Ca}}}}
\newcommand{\gNaF}{\ensuremath{g_{\mathrm{Na,F}}}}
\newcommand{\gKdr}{\ensuremath{g_{\mathrm{K,dr}}}}
\newcommand{\gKA}{\ensuremath{g_{\mathrm{K,A}}}}
\newcommand{\gKD}{\ensuremath{g_{\mathrm{K,D}}}}
\newcommand{\gm}{\ensuremath{g_{\mathrm{m}}}}
\newcommand{\gNaB}{\ensuremath{g_{\mathrm{Na,B}}}}
\newcommand{\gCaB}{\ensuremath{g_{\mathrm{Ca,B}}}}
\usepackage[normalem]{ulem}

\usepackage{scrextend}
\changefontsizes{12pt}

 \journalname{Journal of computational neuroscience}

\begin{document}

\title{Modeling the differentiation of A- and C-type baroreceptor firing patterns}
\thanks{Sturdy and Olufsen were supported in part by the National Science Foundation NSF-DMS \#1246991 and
Olufsen was supported by the National Science Foundation NSF-DMS \#1022688 and \#1122424, by the National Institute of Health NIH-NIGMS \#1P50GM09 4503-01A0, and by the Willumsen Foundation.}

\titlerunning{Modeling the differentiation of A- and C-type neurons}        

\author{Jacob Sturdy \and
            Johnny T. Ottesen \and
            Mette S. Olufsen 
}

\authorrunning{Sturdy, Ottesen, Olufsen} 

\institute{J. Sturdy\at
              Department of Structural Engineering \\
              NTNU \\
              Richard Birkelandsvei 1A\\
              7491 Trondheim, Norway \\
              \email{jacob.t.sturdy@ntnu.no}           
              \and
	     J.T. Ottesen \at
             Department of Science, System and Models\\
             Roskilde University\\
             Universitetsvej 1\\
             4000 Roskilde, Denmark\\
              \email{johnny@ruc.dk}  
	    \and
            M.S. Olufsen \at
            Department of Mathematics\\
            North Carolina State University\\
	   Campus Box 8205\\
	   Raleigh, NC  27695-8205\\
	   Tel: +1 (919) 515 2678\\
	   Fax: +1 (919) 513 7336\\
	   \email{msolufse@ncsu.edu}
}

\date{Received: date / Accepted: date}

\maketitle

\begin{abstract}
The baroreceptor neurons serve as the primary transducers of blood pressure for the autonomic nervous system and are thus critical in enabling the body to respond effectively to  changes in blood pressure. These neurons can be separated into two types (A and C) based on the myelination of their axons and their distinct firing patterns elicited in response to specific pressure stimuli.  
This study has developed a comprehensive model of the afferent baroreceptor discharge built on physiological knowledge of arterial wall mechanics, firing rate responses to controlled pressure stimuli, and ion channel dynamics within the baroreceptor neurons.  With this model, we were able to predict firing rates observed in previously published experiments in both A- and C-type neurons.  These results were obtained by adjusting model parameters determining the maximal ion-chan\-nel conductances.  The observed variation in the model parameters are hypothesized to correspond to physiological differences between A- and C-type neurons. In agreement with published experimental observations, our simulations suggest that  a two\-fold lower potassium conductance in C-type neurons  is responsible for the observed sustained basal firing, where\-as a tenfold higher mechanosensitive conductance is responsible for the greater firing rate observed in A-type neurons. A better understanding of the difference between the two neuron types can potentially be used to gain more insight into the underlying pathophysiology facilitating development of targeted interventions improving baroreflex function in diseased individuals, e.g. in patients with autonomic failure, a syndrome that is difficult to diagnose in terms of its pathophysiology. 

\end{abstract}

\section{Introduction} \label{intro}

The cardiovascular system (CVS) primarily serves to transport substances including oxygen, nutrients, hormones, carbon dioxide, and waste products (\cite{levick_introduction_2010}). The CVS maintains homeostasis via a dominance of negative feedback control, which actively restores the system state in response to perturbations, ensuring an uninterrupted transport function. Inputs encoding the state of the CVS are critical to the system regulation. The baroreceptor neurons monitor blood pressure by sensing changes in arterial wall strain that accompany arterial wall deformation in response to changes in blood pressure. These neurons are  divided into two types according to their myelination and firing rate characteristics: A-type neurons are myelinated and fire at a high frequency when the stimulus reaches a certain  threshold, while C-type neurons are unmyelinated and exhibit  irregular firing, typically at low frequencies~\cite{brown_comparison_1976}. The characteristics considered in this analysis are more representative of autoactive C-type neurons, which fire tonically below threshold pressures. The study by \cite{munch_discharge_1992} notes that between 15\% and 54\% of C-type neurons exhibit tonic firing, yet the origin of these two types of C-type neurons is not well known.

\begin{figure}
\centering
\includegraphics[width=8.5cm]{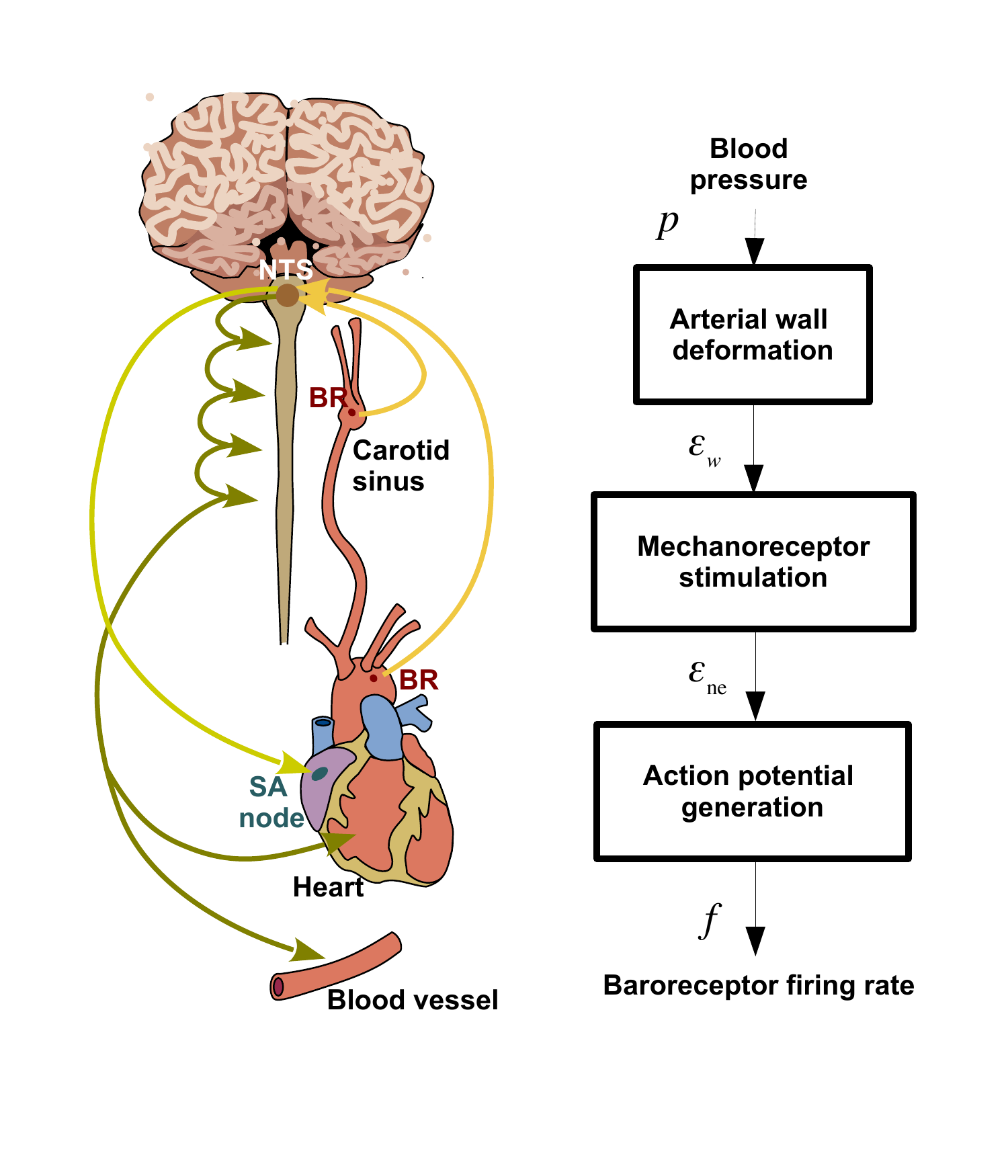}
\vspace{-.75cm}
\caption{The organs and pathways associated with the baroreflex control system are shown on the left. The system's response to a drop in blood pressure includes a decrease in the aortic baoreceptor firing rate, encoding the detected change in arterial wall strain. This signal propagates along afferent baroreceptor fibers to the NTS, which integrates this information into the sympathetic and parasympathetic nervous systems. These in turn increase heart rate, cardiac contractility, and vascular resistance, and decrease vascular compliance. The right panel shows the modeling framework as a block diagram representing the biophysical basis of baroreceptor firing in response to an applied blood pressure stimulus. Abbreviations: BR (baroreceptor nerve endign), NTS (Nucleus Solitary Tract), SA (Sino Atrial node), $p$ (arterial blood pressure), $\epsilon_{w}$ (arterial wall strain), $\epsilon_{\mathrm{ne}}$ (nerve ending strain), and $f$ (baroreceptor firing rate).}
\label{fig:baroreflex}
\end{figure}

As depicted in \fref{fig:baroreflex}, the baroreceptor neurons consist of mechanosensitive sensory nerve endings primarily located in the walls of the aorta and carotid sinuses. These nerve endings connect to dendrites that carry the electrical signal  to the nucleus solitary tract (NTS) (\cite{levick_introduction_2010}), which receives input from the baroreceptor neurons as well as from other cardiovascular afferents, such as chemoreceptors. The NTS integrates these inputs into a combined signal, which is relayed to the areas of the medulla responsible for generating sympathetic, and parasympathetic efferent signals. The targets for the efferent neural signals are the heart,  and the small arteries and arterioles.
At these targets the signals modulate heart rate, cardiac contractility, vascular resistance and compliance to maintain homeostasis (\cite{korner_integrative_1971,levick_introduction_2010,thomas_neural_2011}). 

Beyond observation of differences in the baroreceptor firing rate generated by A- and C-type neurons, \cite{fan_graded_1999} demonstrated that selective stimulation of A- or C-type neurons elicits distinct efferent responses. Both the blood pressure and the heart rate responses are most effective when both types of neurons are stimulated; however, selective stimulation reveals that combined A- and C-type stimulation produces more than twice the change in heart rate compared to A-type or C-type stimulation alone. 
Additionally, selectively stimulating A-type neurons produces a greater maximal change in blood pressure than stimulation of C-type neurons alone, though the frequency of stimulation required to achieve this is much higher than the frequency required to elicit a comparable change through C-type stimulation alone. Though these two results highlight some differences between A- and C-type response, it must be noted that this study also emphasizes how the two populations of neurons contribute to the reflex response across different frequencies of stimulation, with C-type neurons dominating the response to low frequency stimulation, whereas A-type neurons primarily contribute to reflex responses at higher frequencies of stimulation.

Both neuron types are stimulated via activation of mechanosensitive ion channels (MSC) by changes in the wall strain (caused by changes in blood pressure), and thus transduce changes in the blood pressure into an electrically encoded neural signal (\cite{brown_baroreceptor_1978,levick_introduction_2010}). Another observation by~\cite{li_electrophysiological_2008} from electrophysiological and anatomical studies in rats showed that A-type neurons may be separated into two subtypes A- and Ah-type, and that female rats have significantly more Ah-type neurons than male rats. Further studies by~\cite{chavez_afferent_2014} of these fibers have revealed that selective stimulation of myelinated neurons in female rats exhibits a lower threshold for MAP reduction compared to male rats suggesting that Ah-type baroreceptors may provide a functionally distinct afferent pathway within the baroreflex arc.

While many aspects of baroreflex regulation have been studied extensively (\cite{benarroch_arterial_2008}), numerous factors modulating baroreflex responses are not fully understood, e.g. the role of angiotensin in modulation of baroreflex sensitivity (\cite{palma-rigo_angiotensin_2012,saigusa_ANG_2014}), and the mechanical properties facilitating coupling of the nerve endings to the arterial wall (\cite{brown_receptors_1980}). Both of which may be of significance in understanding the role of the baroreflex in hypertension as suggested by the recent study by \cite{pettersen_arterial_2014}.

This is largely due to the difficulty associated with studying baroreceptor function at the cellular level experimentally. In particular, no studies have been able to describe the electrophysiological and mechanical characteristics of nerve endings. The typical approach involves isolating and recording electrophysiological properties in neurons that are separated from their mechanosensitive endings (\cite{kraske_mechanosensitive_1998,snitsarev_mechanosensory_2002}). The nerve endings are best described as a branching and intertwined neural network that is integrated into the adventitial layers of the arterial wall, making it virtually impossible to experimentally isolate  the nerve endings without damage (\cite{kraske_mechanosensitive_1998,krauhs_structure_1979}). 

\begin{figure*}
\centering
\includegraphics[width=\textwidth]{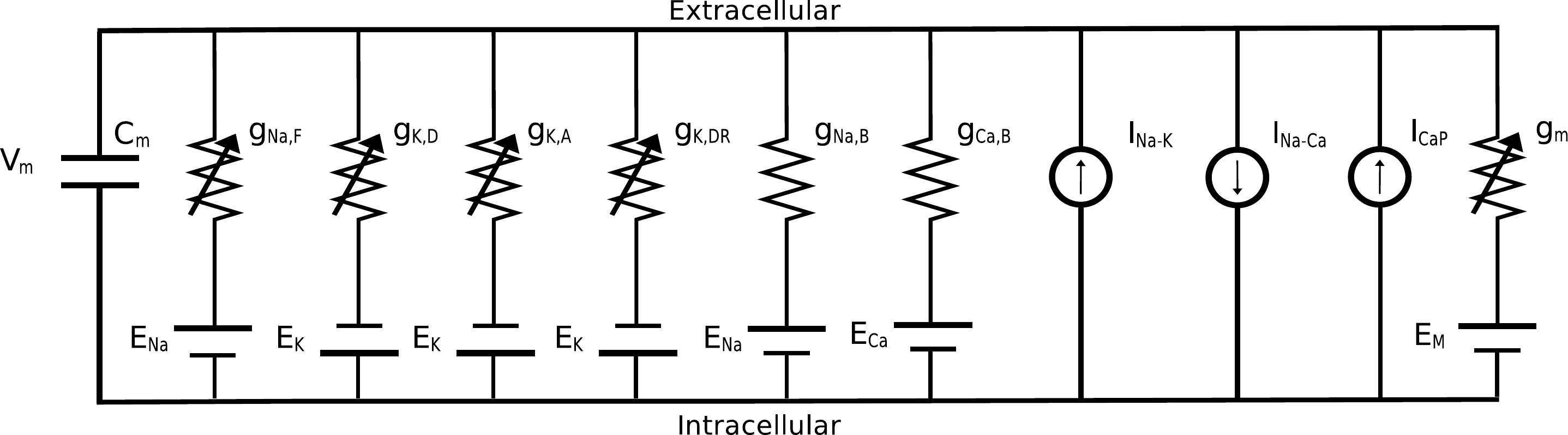}
\caption{ The main channels associated with afferent baroreceptor fibers are a mechanosensitive conductance (\gm{}), a fast sodium conductance (\gNaF{}), 4-AP sensitive potassium conductances (\gKA{} and \gKD{}), a delayed rectifier conductance (\gKdr{}), linear leakage conductances ( \gNaB{} and \gCaB{}),  a sodium-potassium exchanger current (\INaK), a sodium-calcium exchanger (\INaCa), and a calcium pump (\ICaP).}
\label{fig:neuron}
\end{figure*}

Despite this difficulty, studies of the cell membranes in isolated baroreceptor neurons have identified a number of ion channels and have characterized their dynamics. As discussed in a recent review by \cite{schild_differential_2012}, numerous ion-channels have been identified in baroreceptor neurons. This study only includes a subset of these ion-channels (extracted from previous modeling studies by \cite{schild_-_1994} and \cite{li_kca1.1_2011}), as illustrated in \fref{fig:neuron}. Focus was on selecting a subset of channels allowing the model to detect differences between A- and C-type neurons response to pressure stimuli.  The  selected channels include (listed with the mathematical notation representing their current in parentheses): a TTX-sensitive fast sodium current(\INaF{}), a sodium background current (\INaB{}), a calcium background current (\ICaB{}), a sodium-potassium exchanger current (\INaK{}), a sodium-calcium exchanger (\INaCa{}), a calcium pump (\ICaP{}) a delayed rectifier potassium current (\IKdr{}), and a 4-AP (4-aminopyridine) sensitive potassium current (\IKA{} and \IKD{}). Some of the channels excluded are the calcium sensitive potassium current identified by \cite{li_kca1.1_2011} and the TTX insensitive sodium current present in C-type neurons. 
The latter channel allows C-type neurons to 
continue firing when exposed to TTX, while A-type neurons cease firing. This channel would have allowed us to identify another difference between the two neuron types, but given that we do not have data to evaluate this difference, we omitted  this channel. 

Mathematical modeling provides a way to explore the interaction between the ion-channel dynamics and the arterial wall deformation. Several previous modeling studies have investigated baroreceptor dynamics using approaches ranging from simple phenomenological models predicting the firing rate as a function of blood pressure (\cite{spickler_transduction_1968,ottesen_nonlinearity_1997,mahdi_modeling_2013,ottesen_structural_2014}) to biophysical models using a Hodgkin-Huxley type approach to describe the electrical behavior of the isolated neurons (\cite{schild_-_1994}). Some biophysical models (\cite{brederode_experimental_1990,schild_-_1994}) were designed to predict the differences in firing patterns between A-type and C-type neurons; however, these studies did not examine how changes in vessel strain stimulate the stretch-sen\-si\-tive channels. \cite{alfrey_model_1997} accounted for mechanosensitivity, but focused on reproducing A-type firing patterns. In addition, a number of phenomenological models~\cite{coleridge_characteristics_1987, seagard_firing_1990}) describe both A- and C-type firing rates as functions of blood pressure or wall-strain, but do not consider ion-channel dynamics, and thus do not address the basis of differentiation between A- and C-type neurons.

This study aims to combine previous efforts building a fairly simple biophysical model that can distinguish between A- and C-type firing. Our model has potential to help explore system level differences that may be attributed to differences in the distribution or role of A- and C-type neurons, such as gender differences found to play a role in patients with orthostatic intolerance (\cite{santiago_neuropathy_2000}). Orthostatic intolerance describes the inability of an individual's body to effectively regulate changes in blood pressure caused by changes in posture, typically accompanied by frequent syncope episodes. This disorder occurs in five females for every one occurrence in males (\cite{pickering_orthostatic_2002}). We speculate that this may be associated with different ratio of A- vs. Ah-fibers in males and females (see~\cite{li_electrophysiological_2008,chavez_afferent_2014}). 

To study the origin of the firing patterns displayed by A- and C-type neurons. We designed an anatomically and physiologically based model that can predict arterial wall deformation, mechanoreceptor stimulation, and action potential generation. This model is shown to effectively reproduce experimentally observed responses of baroreceptor neurons to various pressure stimuli for both A- and C-type neurons. We discuss what parameters  characterize the two neuron types  and  use the model to simulate C-type neural responses to pressure step and pulse stimuli, which have not been well characterized in previously published experimental studies.

\section{Methods}\label{sec:methods}

\subsection{Summary of experimental data} \label{sec:experimental_data}
Quantitative data describing the  deformation of the rat aortic arch, along with A- and C-type firing characteristics in baroreceptors within the  rat aorta as well as rabbit and canine carotid arteries, were used for analyzing the model developed in this study. 

Data  characterizing arterial wall deformation  were extracted from the experiments by \cite{feng_theoretical_2007}.  These experiments measured the deformation of surgically isolated rat aortic arches in response to a controlled pressure stimulus.  

Baroreceptor firing rate characteristics illustrating the difference between A- and C-type neurons were extracted from experiments in rats, rabbits and dogs (\cite{brown_baroreceptor_1978,franz_small_1971,saum_electrogenic_1976,schild_mathematical_1994,seagard_firing_1990}).  The experimental procedures in these studies were similar. They were all  performed \emph{in situ}  in surgically isolated or excised vessels stretched to their \emph{in vivo} length with the baroreceptor nerve attached.  

To show firing rate characteristics exhibited by baroreceptor neurons (threshold, saturation, overshoot, and adaptation), the vessels were exposed to three prototypical blood pressure stimuli including a continuous ramp increase, a step change, a sinusoidal stimulus, and a pulse pressure stimulus. Data used to validate the model against each stimulus type are described below.

\begin{figure*}[ht!]
\centering
\textbf{A} \includegraphics[width=.45\textwidth]{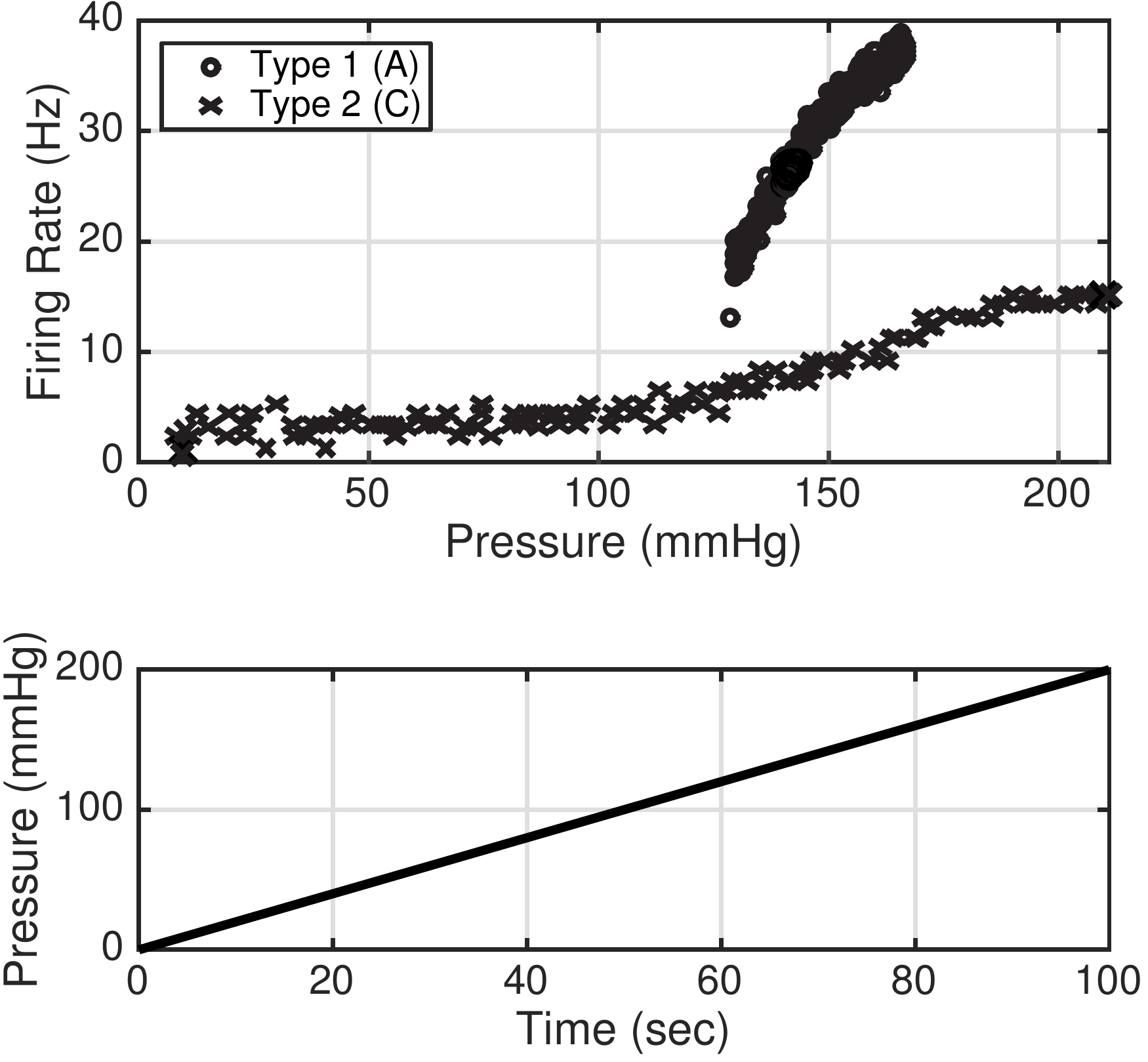}
\textbf{B} \includegraphics[width=.45\textwidth]{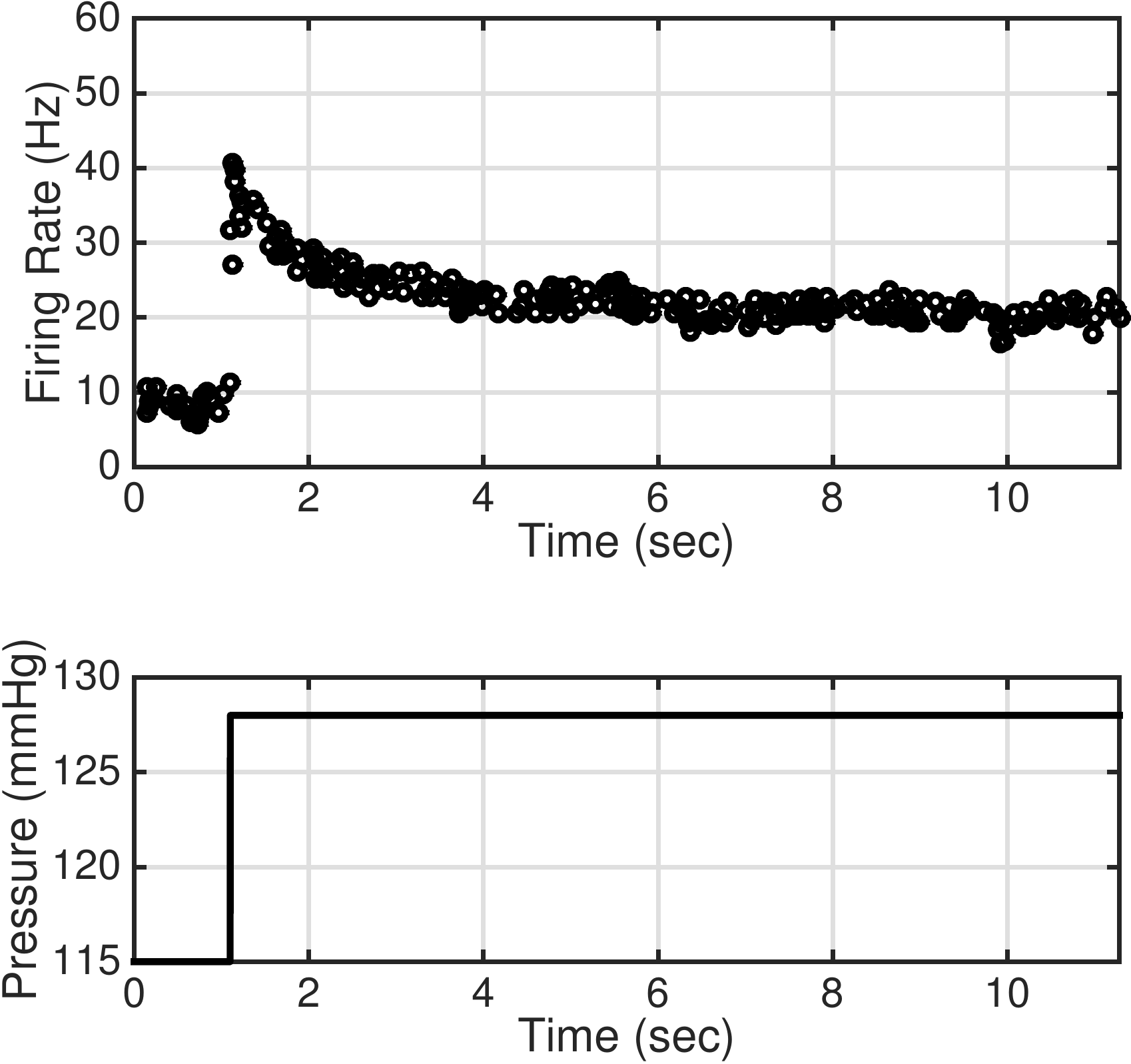}

\textbf{C} \includegraphics[width=.45\textwidth]{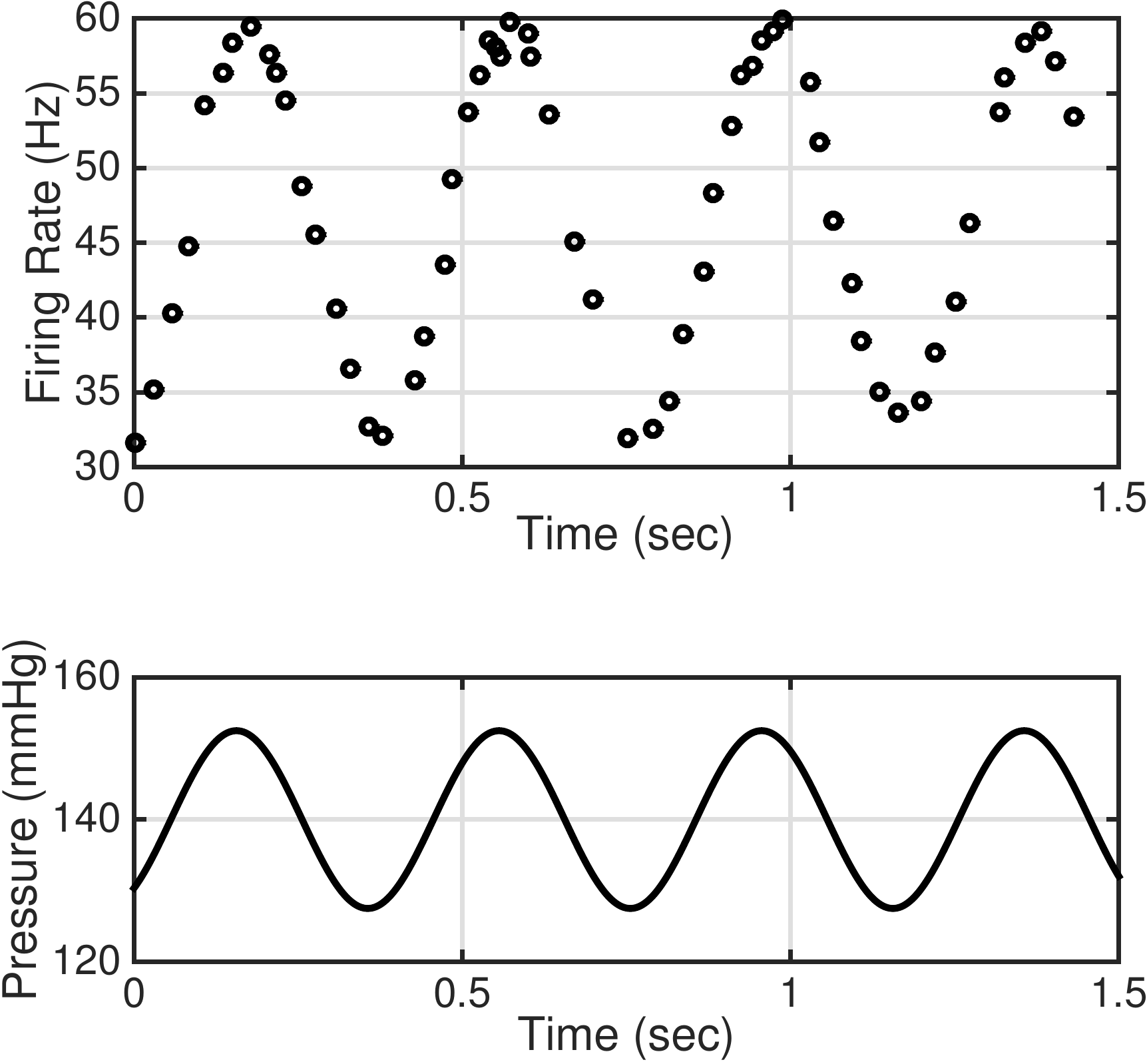}
\textbf{D} \includegraphics[width=.45\textwidth]{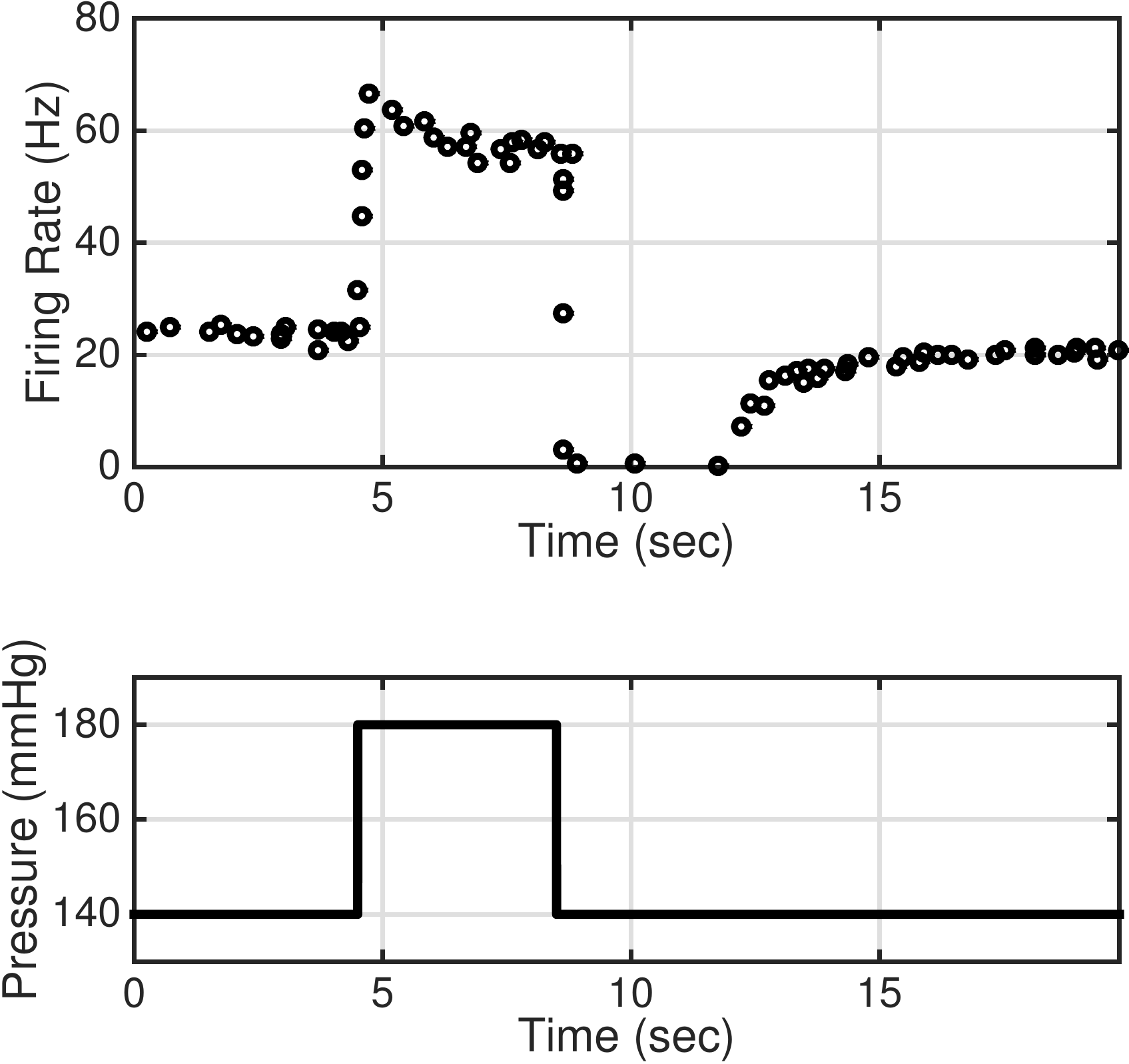}
\caption{\textbf{A} shows a ramp stimulus (bottom) and the associated firing rate response for A- and C-type neurons from a dog carotid artery.  Data are extracted from the studies by \cite{schild_mathematical_1994} and \cite{seagard_firing_1990}. \textbf{B} shows the step response from a rat A-type neuron (it is believed that some C-type neurons respond in a qualitatively similar manner with a lower firing rate, while others exhibit overshoot, but subsequently ceases to fire),  Data extracted from the study by \cite{brown_baroreceptor_1978}.  \textbf{C} shows a sinusoidal stimulus and response for a rabbit A-type neuron, data extracted from \cite{franz_small_1971}. C-type firing patterns are reported to be similar but with lower amplitude (See \cite{brown_baroreceptor_1978}). \textbf{D} shows the pulse pressure stimulus and response for a rat A-type neuron.  Note that the neuron cease firing following the pressure drop.  Data are extracted from \cite{saum_electrogenic_1976}. To our knowledge detailed firing rate recordings of C-type response to a pulse stimulus have been reported.}
\label{fig:data}
\end{figure*}

\paragraph{A ramp stimulus} (see bottom of  \fref{fig:data}A.) can be achieved  using a syringe pump to infuse fluid in a vessel clamped at its outlets. The continuous infusing of fluid cause the pressure to increase at a rate of 1-2 mmHg/sec. Previously published experiments using a ramp stimulus reveal two characteristic features of baroreceptor neurons: threshold and saturation. \textbf{Threshold} characterizes the pressure at which the firing rate frequency suddenly changes.  \fref{fig:data}A shows that A-type neurons start firing when the ramp stimulus reaches a given threshold, whereas autoactive C-type neurons fire at all pressures (over the ramp) but change firing rate at the threshold pressure, \cite{seagard_firing_1990}.\footnote{Seagard et al. refer to A-type as Type I, and C-type as Type II.}  \textbf{Saturation} refers to the pressure range over which the firing rate remains constant, i.e. when a pressure increase no longer leads to an increase in firing rate. While both A- and C-type neurons exhibit saturation, A-type neurons saturate at a higher frequency but lower pressure than C-type neurons (see \fref{fig:data}A.).

A number of studies analyzing the response to a ramp stimulus have achieved similar results (\cite{bolter_relationship_2011,coleridge_characteristics_1987,franz_small_1971,munch_discharge_1992,munch_role_1985,sato_dynamic_1998,spickler_dynamic_1967,tomomatsu_carotid_1983}). Two of these studies (\cite{coleridge_characteristics_1987,spickler_dynamic_1967}) stimulated the baroreceptor neuron with a ramped pulsatile pressure: an underlying increasing mean pressure overlaid with a pulsatile pressure. These studies observed that the pulsatile stimulus shifts the firing rate response up, but does not change the qualitative features.	

In this study, we model the ramp  stimulus using the linear function
\begin{equation}
	p(t) = at + b,
	\label{eq:ramp}
\end{equation}
where $a$ denotes the rate of the blood pressure increase, and $b$ the baseline pressure.  

\paragraph{A step pressure stimulus} is commonly used to characterize the dynamic response of the baroreceptor neurons, and refers to stimulation by  blood pressure changed in a rapid step (up to 200 mmHg over about 100 msec) from one value to another (see \fref{fig:data}B).  \cite{brown_baroreceptor_1978} investigated the rat aortic baroreceptor firing rate in response to four pressure steps increasing pressure from a baseline of 115 mmHg with steps of 13 (to 128), 19 (to 134), 22 (to 137), and 28 (to 143) mmHg (\fref{fig:data}B).  The characteristic response of both A- and C-type neurons to a step stimulus is an \textbf{overshoot followed by adaptation}. At the onset of the pressure increase, neurons dramatically increase their firing rate, after which it  decays toward a new steady firing rate corresponding to the new pressure (see \fref{fig:data}). \cite{kunze_arterial_1991} report that C-type neurons (not shown) exhibit more irregular firing patterns than A-type neurons. Yet~\cite{brown_comparison_1976} report that some C-type neurons respond with an overshoot followed by rapid adaption and then sustained cessation as the step up in pressure is maintained. For both neuron types, the experiments were done over a period of 12 sec, allowing the firing rate to adapt to a new steady level of discharge. Similar responses, mostly in A-type neurons, have been observed in multiple studies (\cite{saum_electrogenic_1976,vliet_response_1987}).

We model the pressure step change using the function
\begin{equation}
	p(t)=
	\begin{cases}
             p_{b} & \text{if } t<t_{\mathrm{step}} \\
             p_{b} + \Delta p & \text{ otherwise,}
	\end{cases}
    \label{eq:step}
\end{equation}
where \(p_b\) denotes the baseline pressure and  \(\Delta p\) the step change.

\paragraph{A sinusoidal pressure stimulus} is typically used to analyze the firing rate dynamics in a setting mimicking \textit{in vivo} conditions. In response, both A- and C-type baroreceptors fire sinusoidally  with some phase shift (\cite{brown_baroreceptor_1978}), though C-type baroreceptors have characteristically lower firing rates  (see \fref{fig:data}C).  Spickler and Kedzi also studied the response of (presumed A-type) baroreceptors and attempted to characterize their frequency response characteristics, finding an increased activity corresponding to increased stimulus fre\-quen\-cy (\cite{spickler_dynamic_1967}).  \cite{franz_small_1971} studied the response of (presumed A-type) baroreceptors in rabbits and attempted to develop a black box input/output model of these based on signal characteristics.  Their sinusoidal stimulus recordings showed similar results to those of \cite{brown_baroreceptor_1978}.

The sinusoidal stimulus is modeled as
\begin{equation}
      p(t) = p_{b} + p_{A} \sin \left(2\pi(\omega t + \phi) \right),
\label{eq:sineStim}
\end{equation}
where  \(p_{b}\) denotes the baseline pressure, \(p_{A}\) the amplitude, and \(\omega\) the stimulus frequency, and $\phi$ specifies the phase shift relative to a single period of the signal.

\paragraph{A pulse pressure stimulus} refers to a step pressure increase followed by a step decrease back to the original pressure level (\fref{fig:data}D). \cite{saum_electrogenic_1976} used this stimulus to investigate the response known as \textbf{Post Excitatory Depression} (PED) observed in A-type neurons. PED is a cessation of baroreceptor firing for a period of up to 10 seconds following the sudden decrease in pressure. The cessation of firing following decreasing pressure has been observed in numerous studies beginning with the one by \cite{bronk_afferent_1932} who noted that baroreceptor neurons cease to fire during diastole.  This phenomenon was analyzed in numerous previous studies: \cite{landgren_excitation_1952} observed effects of amplitude and duration of the pressure step on the duration of PED. \cite{wang_postexcitatory_1991} observed that the PED duration  depends on the duration and the height of the pressure step. Finally,  \cite{saum_electrogenic_1976} demonstrated that PED may be inhibited through processes within the baroreceptor nerve fiber itself. He suggested that an electrogenic sodium pump contributes to the phenomenon. To our knowledge no studies have investigated the response of isolated C-type neurons to a pulse pressure stimulus.
The pressure pulse stimulus is modeled using the function
\begin{equation}
	p(t)=
	\begin{cases}
                p_{b} & \text{if } t<t_{\mathrm{up}} \text{ or } t_{\mathrm{down}}< t \\
                p_{b} + \Delta p & \text{ otherwise,}
	\end{cases}
    \label{eq:pulse}
\end{equation}
where \(p_b\) denotes the baseline pressure and  \(\Delta p\) the step change.

\subsection{Modeling} \label{sec:modeling}
As shown in \fref{fig:baroreflex}, the baroreflex model consists of three components predicting arterial wall deformation, neural deformation and mechanoreceptor stimulation, and action potential generation from which firing rate is extracted. A nonlinear function relating pressure to arterial wall strain is used to predict the increased stiffening observed with an increased pressure stimulus. This model is based on experimental studies in rats (in aortic baroreceptor neurons) (\cite{feng_theoretical_2007,bezie_fibronectin_1998}) and in sheep (in a range of large arteries) (\cite{valdez-jasso_analysis_2009,valdez-jasso_linear_2011}). Changes in wall strain drives the baroreceptor nerve ending deformation. To our knowledge no experimental studies have characterized the coupling of the nerve ending deformation to the wall strain, thus no explicit model validation can be made for this model component. For this part of the model, we incorporated ideas put forward by \cite{alfrey_model_1997}, \cite{bugenhagen_identifying_2010} and \cite{mahdi_modeling_2013} using the assumption that the nerve ending deformation exhibits qualitative dynamics that are similar to that associated with the baroreceptor firing rate response to a step pressure stimulus. For this study we adopt a linear viscoelastic model with two relaxation time scales of approximately 1 and 3 seconds to predict the  nerve ending deformation induced by the arterial wall strain. Nerve ending deformation stimulates mechanosensitive ion channels, whose probability of opening is modeled by a sigmoidal function of the nerve ending strain. Finally, the afferent baroreceptor firing rate is calculated from the action potentials generated by a Hodgkin-Huxley type model incorporating the major ion channels identified in patch clamp studies of baroreceptor cell bodies (\cite{schild_-_1994}).

\subsubsection{Arterial wall deformation}
The large arteries,  have connective tissue attaching them to the surrounding tissues and structures of the body. As a result, the vessels are pre-stretched in their longitudinal direction and therefore mainly deform axially (\cite{fung_biomechanics_1996}). Several recent studies in large mammals, e.g.\ the sheep study by \cite{valdez-jasso_linear_2011}, have shown that arterial deformation displays nonlinear elastic and viscoelastic properties, yet it is not clear if viscoelasticity plays a major role in arterial wall deformation within small animals (\cite{brown_baroreceptor_1978,boutouyrie_vivo/vitro_1997}).  Due to the focus of this study on data from \cite{brown_comparison_1976,brown_baroreceptor_1978} and \cite{feng_theoretical_2007} in rats, and \cite{seagard_firing_1990} in dogs, we have chosen to describe wall deformation using the elastic nonlinear sigmoidal function proposed by \cite{valdez-jasso_linear_2011}. This model accounts for the high stiffness of the arteries at both high and low pressures following 
\begin{equation}
  A(p)=(A_m-A_0)\frac{p^{\kappa_w}}{\alpha_w^{\kappa_w}+p^{\kappa_w}}+ A_0,
\label{eq:sigmoid}
\end{equation}
where \(A_0\) and \(A_m\) are the unstressed and maximum cross-sectional area,  $\alpha_{w}$ represents the pressure at which $A(p)=\sfrac{A_m}{2}$, and $\kappa_w$ determines the steepness of the response.

Following \cite{fung_biomechanics_1993},  the axisymmetric strain can be defined as
\[
    \epsilon_{w} = \frac{r - r_{0}}{r},
\]
where $r$ and $r_0$ denote the actual and unstressed vessel radii. 

Substituting $r=\sqrt{A/\pi}$ into \eqref{eq:sigmoid} gives
\begin{equation}
  \epsilon_w=1-\sqrt{\frac{(\alpha_w^{\kappa_w}+p^{\kappa_w})}{\alpha_w^{\kappa_w}+R_{A} p^{\kappa_w}}},
  \label{eq:NLW}
\end{equation}
where \( R_{A} = \sfrac{A_{m}}{A_{0}} \).

\subsubsection{Nerve ending deformation and mechanoreceptor stimulation}
The aortic and carotid baroreceptor nerve endings form a complex branching network within the outermost layer (the adventitia) of the arterial wall.  Exiting the adventitia individual axons merge into the afferent vagal nerve (cranial nerve X) (\cite{andresen_nucleus_1994}).  \cite{krauhs_structure_1979} studied the anatomy of the baroreceptor-endings in the arterial walls of rat aortic arches. The nerve endings typically lay in the collagenous tissue between elastic laminae within the adventitia.  They found nerve endings both with and without connective fibers attaching them to the tissues of the arterial wall. It is well known that collagenous tissues display both elastic and viscoelastic deformation (\cite{fung_biomechanics_1993}); however, how these properties determine the transfer of arterial wall strain and stress to the nerve endings have, to our knowledge, neither been measured nor modeled.

\begin{figure}[ht!]
\centering
\includegraphics[width=.4\textwidth]{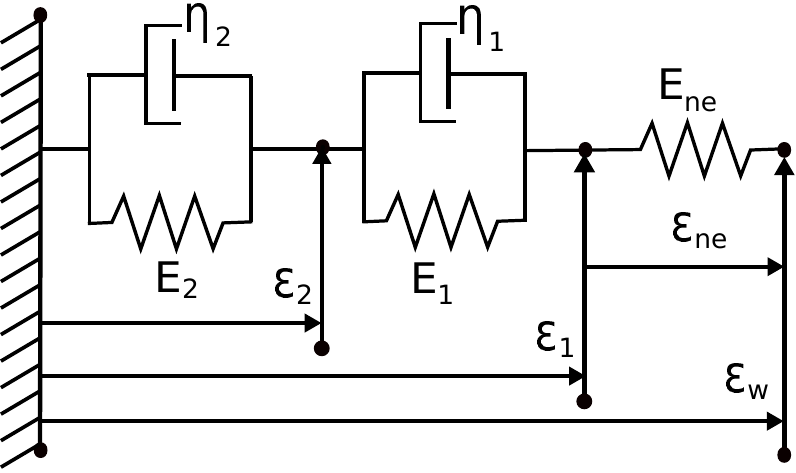}
\caption{Linear mechanical model predicting the transfer of wall strain to the nerve ending strain. The strain across the coupling system is assumed to equal the circumferential wall strain, $\epsilon_{w}$. The strain experienced by the spring labeled $E_{ne}$ corresponds to the strain transferred from the wall to the nerve endings, $\epsilon_{ne }$.}
\label{fig:V2Coupling}
\end{figure}

In this study we model the deformation following ideas by \cite{alfrey_model_1997}, \cite{bugenhagen_identifying_2010}, and \cite{mahdi_modeling_2013} that described the displacement of the nerve endings relative to the total wall stretch using a model with two Voigt bodies in series with a spring. These are all in parallel with the wall deformation, \( \epsilon_{w} \), as presented in \fref{fig:V2Coupling}.

Springs have the stress-strain relation $\sigma =  E \epsilon$, while the dash-pot elements follow $\sigma = \eta \der{\epsilon}{t}$. The strain across elements in parallel is equal, while the stress of elements in series are equal. Applying these relations to the first Voigt body gives the stress-strain equation \(\sigma_{2} = E_{2}\epsilon_{2} + \eta_{2} \der{\epsilon_{2}}{t} \), where $\epsilon_{2}$ is the strain across the Voigt body. The second Voigt body has a total strain \((\epsilon_{1} - \epsilon_{w})\) giving \(\sigma_{1} = E_{1}(\epsilon_{1} -\epsilon_{2}) + \eta_2\der{}{t}(\epsilon_{1} -\epsilon_{2})\). Both \(\sigma_1\) and \(\sigma_{2} \) must be equal to the stress in the final spring element \(\sigma_{ne} = E_{ne}(\epsilon_{w} - \epsilon_{1})\). By substituting the derived expressions in \(\sigma_{2} = \sigma_{ne}\) gives the following differential equation for \(\epsilon_{2}\)
\begin{equation*}
   \der{\epsilon_{2}}{t} =  -\frac{E_2}{\eta_{2}} \epsilon_{2}  + \frac{E_{ne}}{\eta_{2}}(\epsilon_{w} - \epsilon_{1}).
\end{equation*}
Similarly solving \(\sigma_{1}=\sigma_{ne}\) yields
\begin{equation*}
\begin{aligned}
    \der{\epsilon_{1}}{t} = & -\left(\frac{E_{ne}}{\eta_{1}} + \frac{E_{ne}}{\eta_{2}} + \frac{E_{1}}{\eta_{1}}\right)\epsilon_{1}
    + \left( \frac{E_{1}}{\eta_{1}}- \frac{E_{2}}{\eta_{2}}\right)\epsilon_{2} \\
    & +  \left( \frac{E_{ne}}{\eta_{1}}- \frac{E_{ne}}{\eta_{2}}\right)\epsilon_{w}.
\end{aligned}
\end{equation*}
Further details explaining this model can be found in the study by \cite{mahdi_modeling_2013}.

One parameter can be eliminated by re-pa\-ra\-me\-te\-ri\-za\-tion of this system using \(\beta_{i} = E_{i}/\eta_{i}\)  and \( \alpha_{i}= E_{ne}/\eta_{i}\) giving
\begin{eqnarray}
\frac{d\epsilon_1}{dt}&=&-(\alpha_1\!+\!\alpha_2\!+\!\beta_1)\epsilon_1 \!+\! (\beta_1\!-\!\beta_2)\epsilon_2\!+\! (\alpha_1\!+\!\alpha_2)\epsilon_{w}\nonumber \\
\frac{d\epsilon_2}{dt}&=&-\alpha_2\epsilon_1 - \beta_2\epsilon_2  + \alpha_{2} \epsilon_{w}, \label{eq:V2Coupling} 
\end{eqnarray}
where, the total strain \(\epsilon_{ne}\) experienced  by the mechanoreceptor is given by
\begin{equation}
    \epsilon_{ne} = \epsilon_w -\epsilon_1.
\label{NEV2}
\end{equation}

\paragraph{Mechanosensitive ion channels:}
Previous studies (\cite{sharma_mechanical_1995,kraske_mechanosensitive_1998,cunningham_mechanical_1997,drummond_molecular_1998,snitsarev_mechanosensory_2002}) have identified a mechanosensitive current in baroreceptor cells; however, the details of its activation and voltage-current relationship have not been fully characterized.  We described the current using  a simple Ohmic relation with a reversal potential $E_{\mathrm{m}}$, following the study by (\cite{alfrey_model_1997}). The current is thus
\begin{equation}
      \IM = p_o(\epsilon_\mathrm{ne}) \gm (V-E_{\mathrm{m}}),
\end{equation}
where \gm{} represents the maximal whole cell conductance and $p_{o}(\epsilon_{\mathrm{ne}})$  the fraction of channels open for a given strain.

Following observations by \cite[Figure 2B]{kraske_mechanosensitive_1998}, we assume that the fraction of open channels  $p_o$ depends sigmoidally on the nerve ending deformation, $\epsilon_{\mathrm{ne}}$. This is quantified using a Boltzmann relationship
\begin{equation}
\label{eq:po}
     p_{o}(\epsilon_{ne}) = \left\{ 1 + \exp\left(\frac{\epsilon_{\nicefrac{1}{2}} - \epsilon_{\mathrm{ne}}}{S_{\nicefrac{1}{2}}}\right)\right\}^{-1},
\end{equation}
where \( S_{\nicefrac{1}{2}} \) determines the steepness of the transition, and \(\epsilon_{\nicefrac{1}{2}} \) corresponds to the strain associated with 50\% of the channels in the open state. This basic approximation assumes instantaneous dynamics, which is reasonable since the dynamics of this current is thought to be fast compared to the duration of the action potential (\cite{kraske_mechanosensitive_1998}).

\subsubsection{Afferent action potential generation model}

The spiking activity of the nerve ending is modeled using a simplified conductance based approach predicting action potential generation using voltage-gated channel dynamics. This model uses a single compartment to represent the spike initiation zone and assumes that the generated spikes are carried by the nerve fibers without further modulation. The approach presented here follows the previous study by \cite{schild_-_1994}, though it includes only the following ion channels (see also \fref{fig:neuron}):
\vspace{-.25cm}
\begin{itemize}
\item[$\bullet$] Mechanosensitive current, \IM{} (stretch sensitive, inward).\\[-.75cm]
\item[$\bullet$] TTX-sensitive fast sodium current, \INaF{} (voltage-gated, inward).\\[-.75cm]
\item[$\bullet$] Sodium background current, \INaB{} (inward).\\[-.75cm]
\item[$\bullet$] Calcium background current, \ICaB{} (inward).\\[-.75cm]
\item[$\bullet$] Sodium-potassium pump current, \INaK{} (outward).\\[-.75cm]
\item[$\bullet$] Calcium pump current, \ICaP{} (outward).\\[-.75cm]
\item[$\bullet$] Sodium-Calcium exchanger current, \INaCa{} \ (outward).\\[-.75cm]
\item[$\bullet$] Delayed rectifier current, \IKdr{} \ (voltage-gated, outward).\\[-.75cm]
\item[$\bullet$] 4-AP sensitive potassium currents, \IKA{}+\IKD{} (consists of two independent voltage-gated currents, outward).\\[-.75cm]
\end{itemize}
These channels are chosen due to their relatively large maximal conductances as well as their importance in producing the known qualitative dynamics associated with baroreceptor firing. A Hodg\-kin-Huxley type neuron model is used to describe the voltage of the nerve ending (\cite{koch_methods_1998,izhikevich_dynamical_2007}) formulated using an equivalent circuit with a capacitor in series with conductance pathways representing specific ways current can flow through the membrane of the nerve fiber.
The transmembrane voltage-potential \(V\) is modeled as
\[
  \frac{dV}{dt} = -\sfrac{\sum_{i} I_{i}}{C_m},
\]
where \( I_{i} \) corresponds to the total current through a particular ion channel (including the mechanosensitive channel), and \(C_m \) denotes the membrane capacitance. An equivalent circuit representation of the model is shown in \fref{fig:neuron}.

For each channel, the current through the channel is given by
\begin{equation}
   I_{i} = g_{i} a^k b^l (V-E_{i}) ,
   \label{eq:current}
\end{equation}
where $g_{i}$ denotes the maximum whole-
 in the absence of inactivation, $V$ is the membrane potential, $E_{i}$ is the reversal potential, while $a$ and $b$ denote the activation and inactivation gating variables, where $k$ and $l$ are constants. Each of the gating variables can attain a value between 1 (fully permeable to ions) and 0 (fully non-permeable). The product of these variables denotes the percentage of conducting channels. The integer power, $k$ or $l$, denotes the number of gating particles which must transition in order for the channel to open or close. Assuming the particles are independent, the probability that $k$ activating and $l$ inactivating particles exist in the permeable state is $a^k b^l$.

The dynamics of the ion channel are determined by the gating variables, \(a\) and \(b\), and are modeled by
\[
  \frac{dz}{dt} = \frac{z_{\infty} - z}{\tau_z},
\]
where $z$ represents the gating variable ($a$ or $b$), $z_\infty$ is the steady state value, and $\tau_z$  the characteristic time scale.

For most currents, $z_\infty$ is assumed to exhibit a sigmoidal voltage-dependency
\[
  z_\infty = \left\{1 + \exp\left({\frac{V_{\nicefrac{1}{2}} - V}{S_{\nicefrac{1}{2}}}}\right)\right\}^{-1},
\]
where $V_{\nicefrac{1}{2}}$ corresponds to half-activation potential and $S_{\nicefrac{1}{2}}$ is related to the reciprocal of the slope of the activation curve measured at $V=V_{\nicefrac{1}{2}}$. The time constant $\tau$ (also voltage-dependent) follows a simple Gaussian form
\[
	  \tau = A \exp[-B^2(V-V_{peak})^2] + C,
\]
where $A$ corresponds to the peak amplitude, $B$  scales the function width, and $V_{\mathrm{peak}}$ corresponds to the membrane potential at which $\tau$ equals $A+C$.

The key state variables regulating channel opening and the parameters determining the total Ohmic current are given in Appendix~\ref{sec:appendix}.  For a more thorough treatment of neuronal modeling, we suggest the works of \cite{koch_methods_1998} and \cite{izhikevich_dynamical_2007}.

\subsubsection{Firing rate calculation}
The euqations presented so far described the voltage of the neuron as a continuous function of time; however, the data of interest is the firing rate measured from the timing of spikes in the originally recorded data~\cite{brown_comparison_1976}. An algorithm to automatically calculate the firing rate from the voltage trace described by the model allows efficient comparison of the model voltage trace to the firing rate data.
The algorithm first identifies the timing of action potential spikes, and then calculates the instantaneous firing rate $f$ as the reciprocal of the inter-spike interval. 

To compute action potential timing, the algorithm first detects times, $t_j$, when the voltage rises above a threshold voltage, $V_{\mathrm{ref}}$. It then iterates through the solution data points $(V_i,t_i)$ and identifies a crossing if 
$(V_i-V_{\mathrm{ref}})(V_{i-1}-V_{\mathrm{ref}}) < 0$ and $V_i > V_{\mathrm{ref}}$. If this condition is met then $j$ is incremented starting from 0, and the crossing time $t_j$ is calculated as
\begin{equation}
t_j = V_{\mathrm{ref}} - V_{i-1}\Delta t /(V_{i}-V_{i-1}) + t_i - \Delta t,
\end{equation}
where $\Delta t= t_i - t_{i-1}$.
$ V_{\mathrm{ref}}$ was set to 40 mV. 
The time between consecutive crossings, \(T_{j} = t_{j} - t_{j-1}\), is used to calculate the frequency \(f_{j} = 1/T_{j}\).  To determine if the neuron has ceased firing, a threshold $T_{\mathrm{max}}$ is set. If $T_{\mathrm{max}}$ milliseconds pass between successive action potentials, the instantaneous firing rate is set to  $0$ Hz (We define $T_{\mathrm{max}} = 300 \ \mathrm{msec}$). Finally, piece-wise linear interpolation is used to obtain continuous firing rate $f$.

\subsection{Parameter Estimation and Curve Fitting}
Model parameters are estimated using a combination of hand tuning and automated estimation, minimizing the least squares cost
\begin{equation}
   J(\mathbf{\theta}) = \sum_{i}^{N} \{ y_{d}(t_i) - y_{m}(t_i,\mathbf{\theta})\}^{2},
\label{eq:lsqObj}
\end{equation}
where  \(\mathbf{\theta}\) denotes the vector of model parameters, \( y_{d}(t_{i})\) the data measured at  time \(t_{i}\), and \(y_{m}(t_{i},\mathbf{\theta})\) the associated model values. Automated parameter estimation methods include the Levenberg-Marquardt method and the Nelder-Mead simplex algorithm. The Matlab (\cite{matlab_version_2015}) function \emph{lsqnonlin} (implementing the Le\-ven\-berg-Mar\-quardt method) was used to estimate parameters of the arterial wall model.  To remedy the relatively low speed available in Matlab, for estimation of neuronal parameters and we used the Nelder-Mead optimization tool within the \emph{JSIM} modeling environment (\cite{butterworth_jsim_2014}).  

To compare model behavior to available experimental data and known qualitative features, the individual component models were first calibrated independently. Data for arterial wall deformation recorded by \cite{feng_theoretical_2007} was used to calibrate the arterial wall deformation model. The nerve ending deformation model used parameter estimates reported in a previous study that fitted this model to step response recordings of baroreceptor firing rates (\cite{mahdi_modeling_2013}). The neuronal model was adjusted to reproduce the minimal and maximal firing rates present in data. Subsequently, adjustments were made to better fit the complete model to a particular experimental data with a given stimulus: ramp, sine, step and pulse pressure.

\section{Results}

\subsection{Baseline model calibration}

\begin{figure}[th!]
\centering
\includegraphics[width=.4\textwidth]{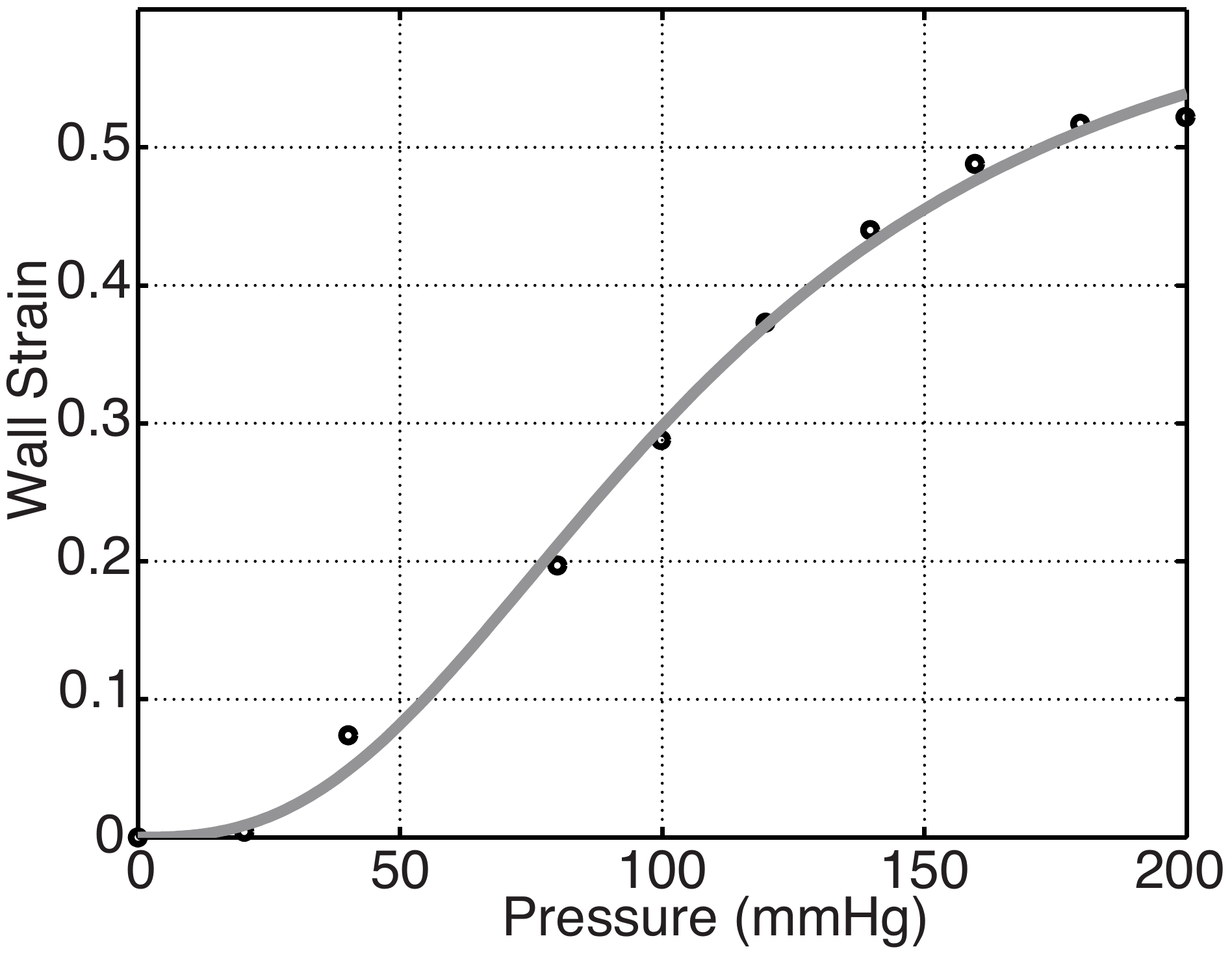}
\caption{Optimized wall strain~\eqref{eq:NLW} as a function of pressure along with aortic deformation data from a rat extracted from \cite{feng_theoretical_2007}. }
\label{fig:WallFitsFeng}
\end{figure}

\paragraph{Arterial wall deformation} predicted by~\eqref{eq:NLW} was compared to data extracted from studies in rat aortic baroreceptors by \cite{feng_theoretical_2007}  (shown in \fref{fig:WallFitsFeng}). This data displays wall deformation in the center of the aortic arch over a range of pressures from $ 0-200$ mmHg~\cite[Figure 4B]{feng_theoretical_2007}. Results shown in \fref{fig:WallFitsFeng} were obtained using the  Matlab function \emph{lsqnonlin} to estimate parameters in~\eqref{eq:NLW} minimizing the least squares cost~\eqref{eq:lsqObj}. Estimated parameter values are $ R_{A} = 8.32$, $\alpha_{w} = 198$ and  $\kappa_{w}=2.65$.

\paragraph{Nerve ending deformation}  was calibrated to produce a response to step pressure changes that qualitatively mimicked the baroreceptor firing rate. We used parameters reported in the previous study by \cite{mahdi_modeling_2013}. This study used a simple model to scale \(\epsilon_{\mathrm{ne}}\) directly to the nerve ending firing rate using an affine function of the nerve ending strain of the form
\begin{equation}
     f_{af}(\epsilon_{ne}) = s_{1} \epsilon_{ne} + s_{2}.
\label{eq:affineMod}
\end{equation}
Reported parameters are $s_1=480$, $s_2=100$, $\alpha_{1}=5.0\cdot 10^{-4}$, $\alpha_{1}=5.0\cdot 10^{-4}$, $\alpha_{2}= 4.0\cdot 10^{-4} $, $ \beta_{1}=5.0\cdot 10^{-4}$, and $ \beta_{2}=2.0\cdot 10^{-3}$ (see \eqref{eq:V2Coupling} for more details).

\paragraph{Neural model} parameters were initially set to values reported by \cite{schild_-_1994}. Parameters for the mechanosensitive channel, $\epsilon_{\nicefrac{1}{2}}$ and $S_{\nicefrac{1}{2}}$, were adjusted to obtain a sigmoidal relattionship of $p_{o}(\epsilon_{\mathrm{ne}})$ over the range of input pressures $p$ pre\-sent in the data ($0-200$  mmHg). $\epsilon_{\nicefrac{1}{2}}$ was set to correspond to the neuron strain achieved at middle pressure between threshold and saturation. The $S_{\nicefrac{1}{2}}$ value was chosen similarly ensuring that $p_{o}$ was nearly zero at the pressure threshold. Finally the value for $\bar{g}_{m}$ was chosen by finding the minimum conductance required to initiate continuous firing in the neuronal model. The resulting values used are presented in~\tref{tab:ParsDesc} in Appendix~\ref{sec:appendix}.

\begin{figure}
  \centering
  \includegraphics[width=0.4\textwidth]{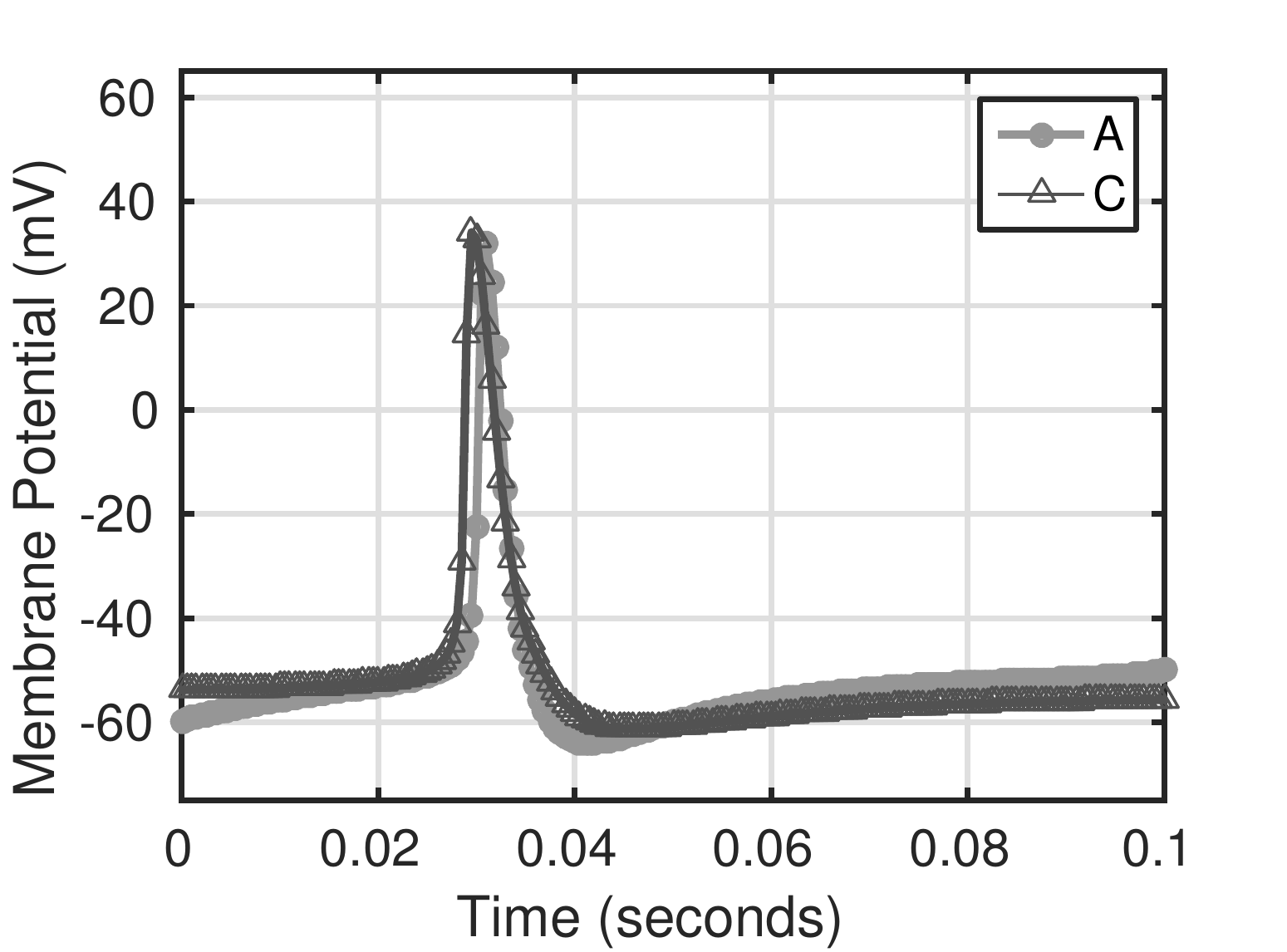}
  \caption{Example voltage traces for model simulations using A-type parameters (circles) and C-type parameters (triangles). The A-type action potential is narrower, while the C-type action potential has a  slightly wider wave form in agreement with observations reported by \cite{schild_-_1994}.}
  \label{fig:ActionPotentials}
\end{figure}

Two model simulated action potentials are depicted in \fref{fig:ActionPotentials}. The figure shows voltage traces from simulations with a constant pressure stimulus for an A-type (circles) and C-type (triangles) neuron.

\subsection{Pressure stimuli simulations}
 For each simulation, model parameters were estimated ensuring that the model qualitatively and quantitatively fits literature experimental data. Given these data sets are recorded under different experimental conditions and in different species (e.g. rats, cats, dogs), parameter values differ between simulations. To remedy this deficiency, the resulting parameters estimated for each stimulus were verified to produce qualitatively consistent behavior across stimuli types.

\begin{figure}
\centering
\includegraphics[width=.4\textwidth]{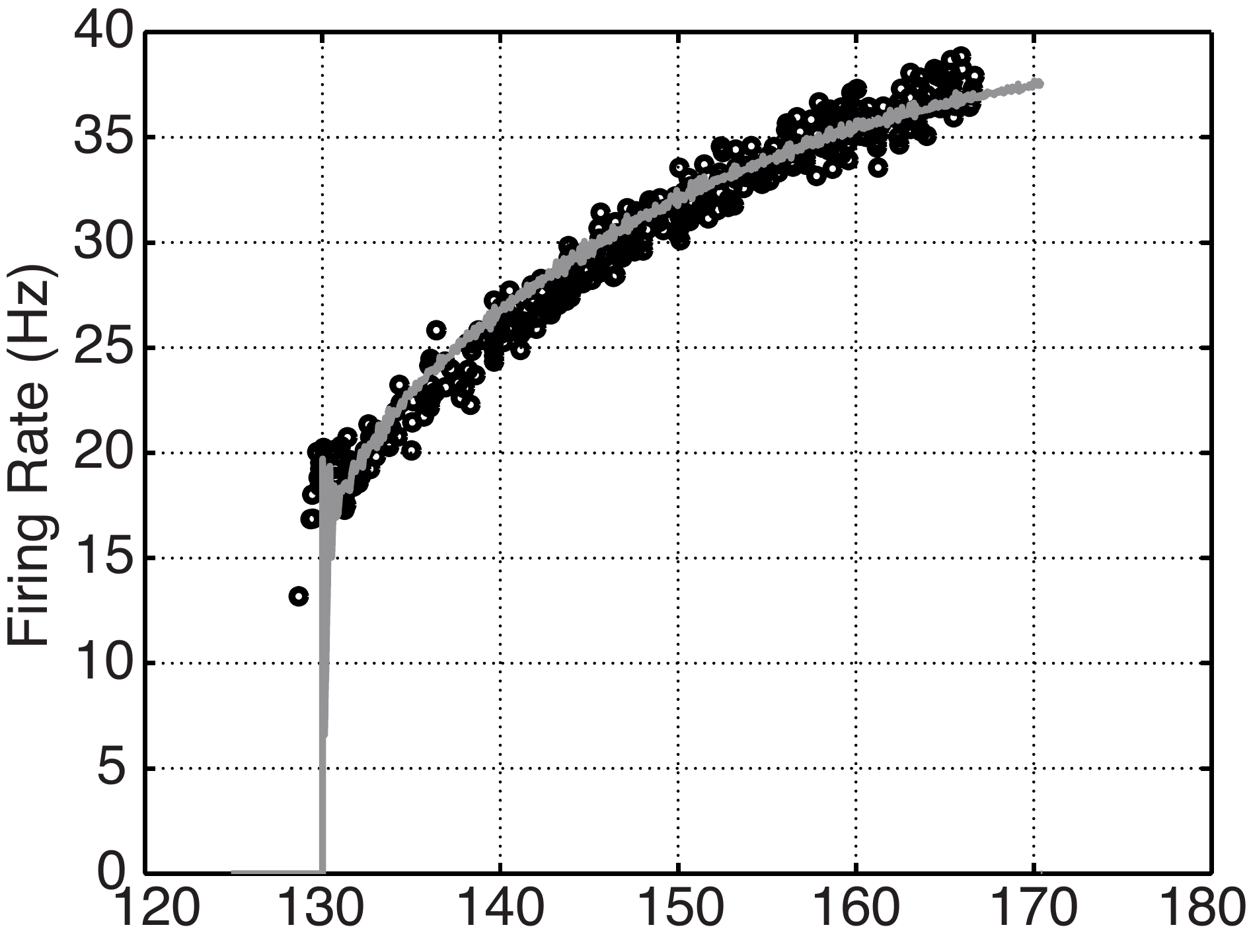}\vspace{.25cm}
\includegraphics[width=.4\textwidth]{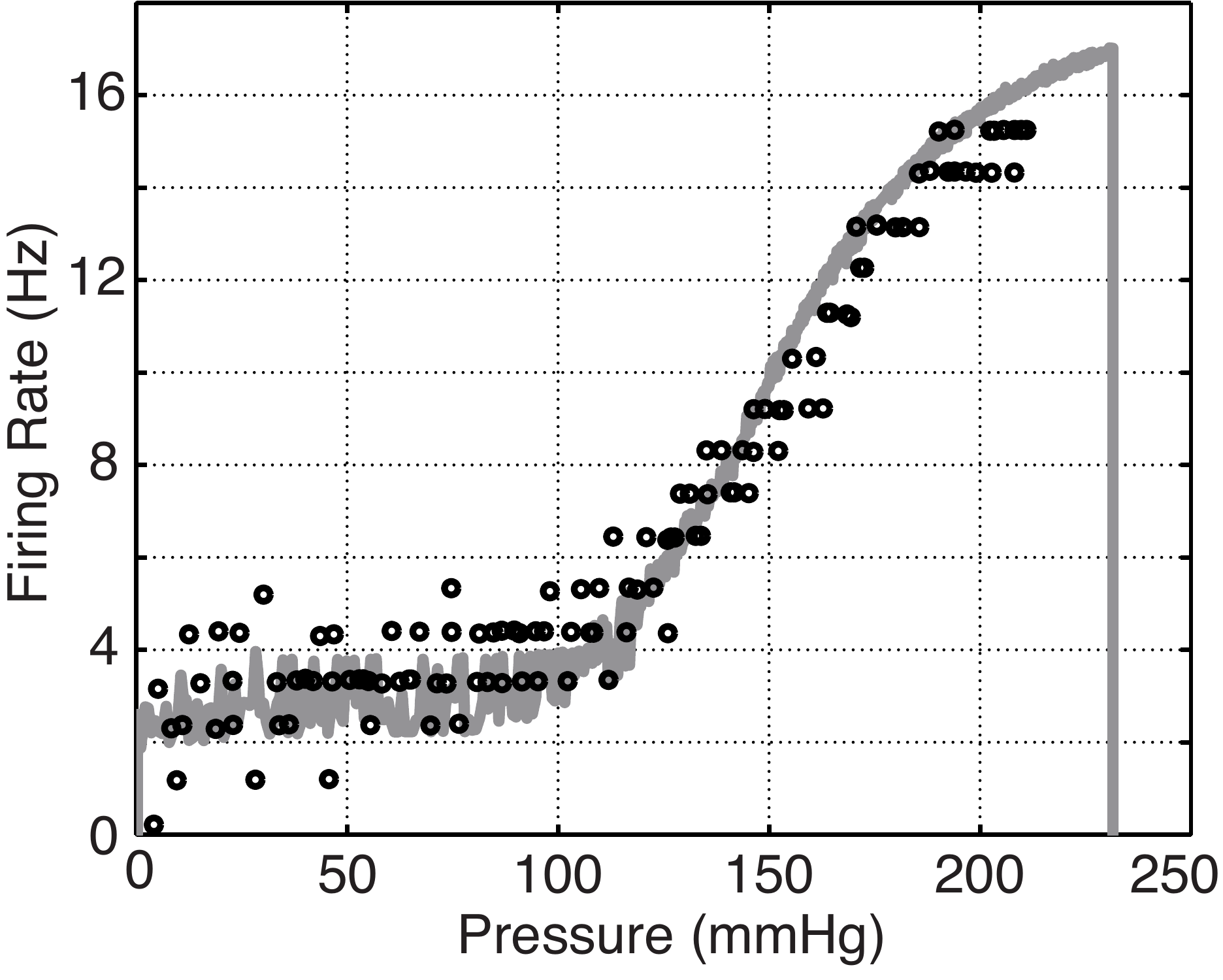}
\caption{Firing rate responses are shown for model simulations with 2 different parameters sets. The parameters used were estimated by minimizing \eqref{eq:lsqObj} comparing the firing rate output to rat A-type data from (\cite[Figure 3.4]{schild_mathematical_1994}) (top), whereas those in the bottom panel were estimated against dog C-type data (Note the C-type data are clustered about integer values which is also the case in (\cite{seagard_firing_1990})). We believe this is not physiological, but a result of either the measurement technique or data post-processing used in the study.}
	\label{fig:RampResults}
\end{figure}

\paragraph{Ramp stimulus:}
To evaluate the model's ability to reproduce both A- and C-type firing rates, we stimulated the model with a ramp pressure  \eqref{eq:ramp} with slope  $a=2$  mmHg/sec. The nerve ending deformation parameters ($\alpha_{1}$, $\alpha_{2}$, $\beta_{1}$, and $\beta_{2}$), half activation strain, $\epsilon_{\nicefrac{1}{2}}$, and reciprocal slope, $S_{\nicefrac{1}{2}}$, as well as the model's max conductances, $g_i$, were estimated using the Nelder-Mead method minimizing \eqref{eq:lsqObj} between model firing rate and the A- or C-type data sets shown in \fref{fig:data}. The initial pressure step parameters were $p_{b} = 100$ mmHg for the A-type simulation and $p_{b} = 0$ mmHg for the C-type simulation.  \fref{fig:RampResults} shows the resulting firing rates~\tref{tab:ParsResults}.

\begin{figure}
\centering
\includegraphics[width=.4\textwidth]{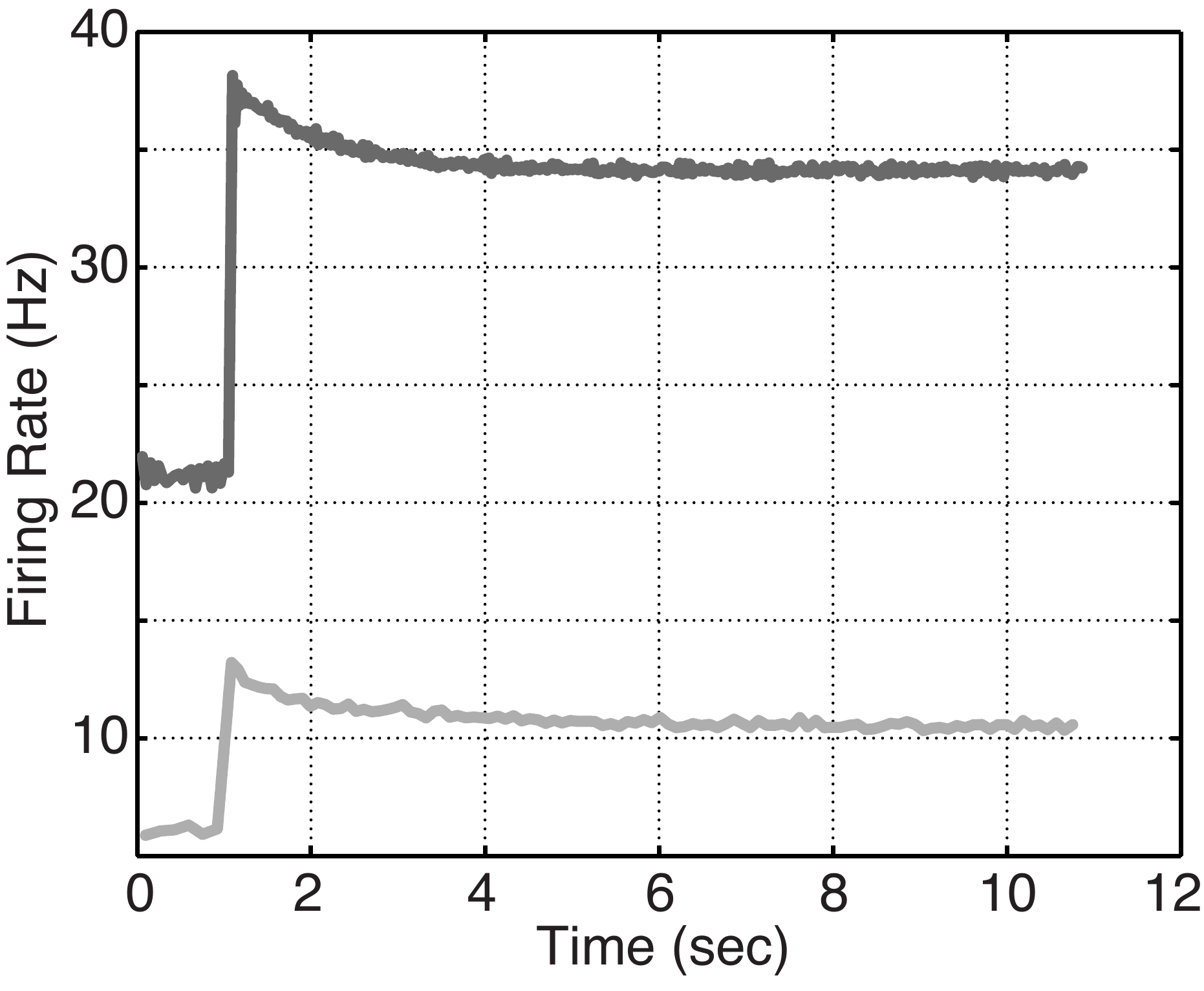}
\caption{Model response to a step pressure input using the model parameters estimated to fit the ramp data: A-type dark, and C-type light.}
\label{fig:rampStepResp}
\end{figure}

In addition to estimating parameters to fit the ramp stimulus, the estimated parameters were used to simulate model response to a step stimulus of the form ~\eqref{eq:step} with base pressure $p_{b} = 135$ mmHg, step amplitude $ \Delta p = 22$ mmHg, and onset time  $t_{\mathrm{step}} = 1.1$ sec. This was done to determine the response of these optimized ramp responses to a step pressure input. The resulting firing rates are shown in \fref{fig:rampStepResp}.

\begin{figure}
\centering
\includegraphics[width=.45\textwidth]{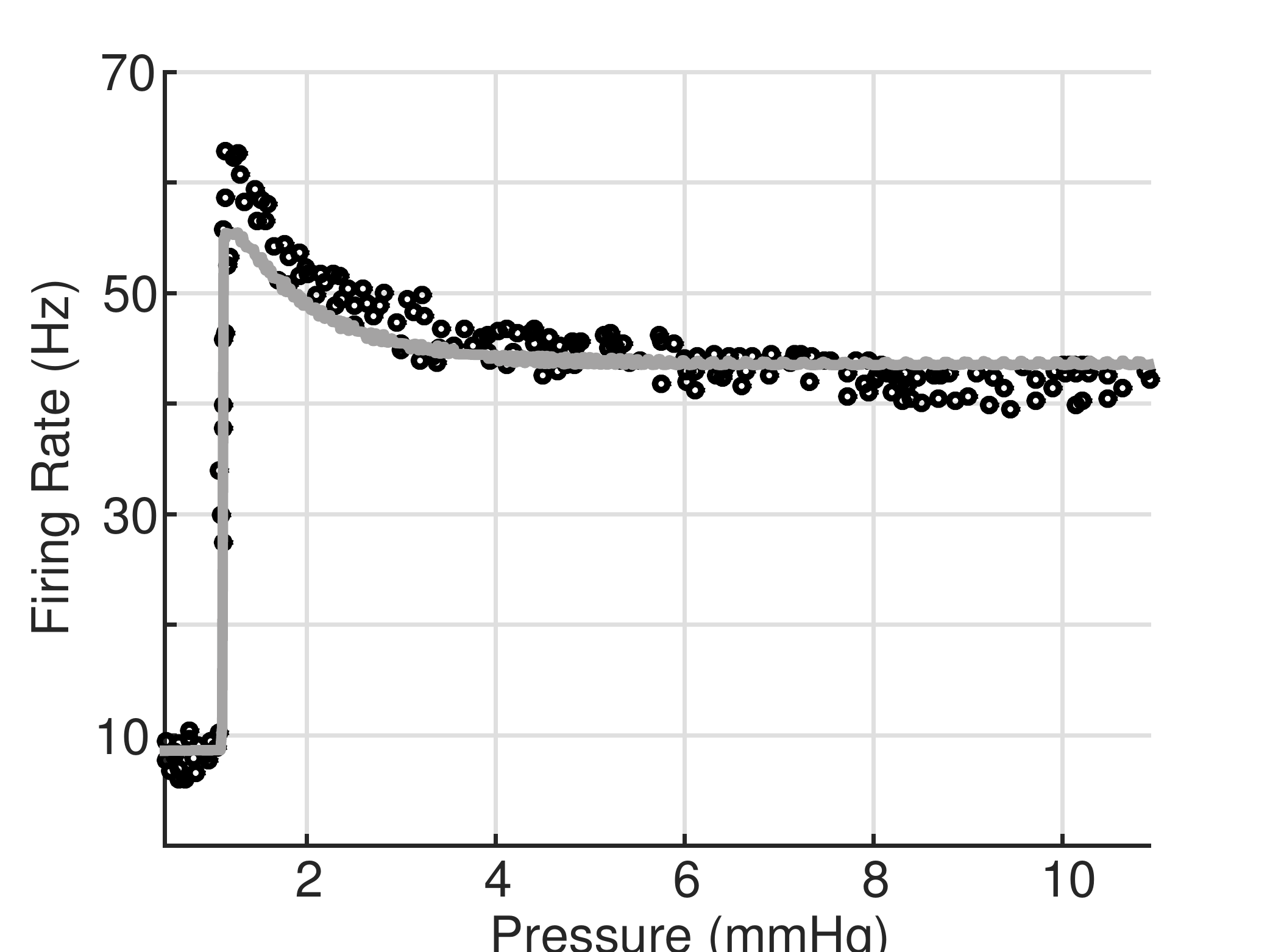}
\caption{Step response for a slowly adapting (rat A-type) neuron observed by \cite{brown_baroreceptor_1978}, and shown in \fref{fig:data}. Estimated parameter values are reported  in~\tref{tab:ParsResults}.}
\label{fig:StepResults}
\end{figure}

\paragraph{Step stimulus:}
A step stimulus~\eqref{eq:step} with $p_{b} = 115$ mmHg, $\Delta p = 22$ mmHg, and $t_{\mathrm{step}} = 1.1$ sec was used to simulate the experiments reported by \cite{brown_baroreceptor_1978}. Similarly to the ramp simulations, mechanical coupling parameters and  neuronal conductances were estimated to fit the data (see \fref{fig:StepResults} and~\tref{tab:ParsResults}). The estimated parameters were also used to simulate the model response to a ramp stimulus with a baseline pressure $p_{b} =0$ mmHg/sec and a slope $a = 2$ mmHg/sec (the values used in the previous simulations). This was done to ensure the parameters estimated for the step response also reflected an appropriate ramp response (results not shown).

\begin{figure*}
\centering
\includegraphics[scale=.5]{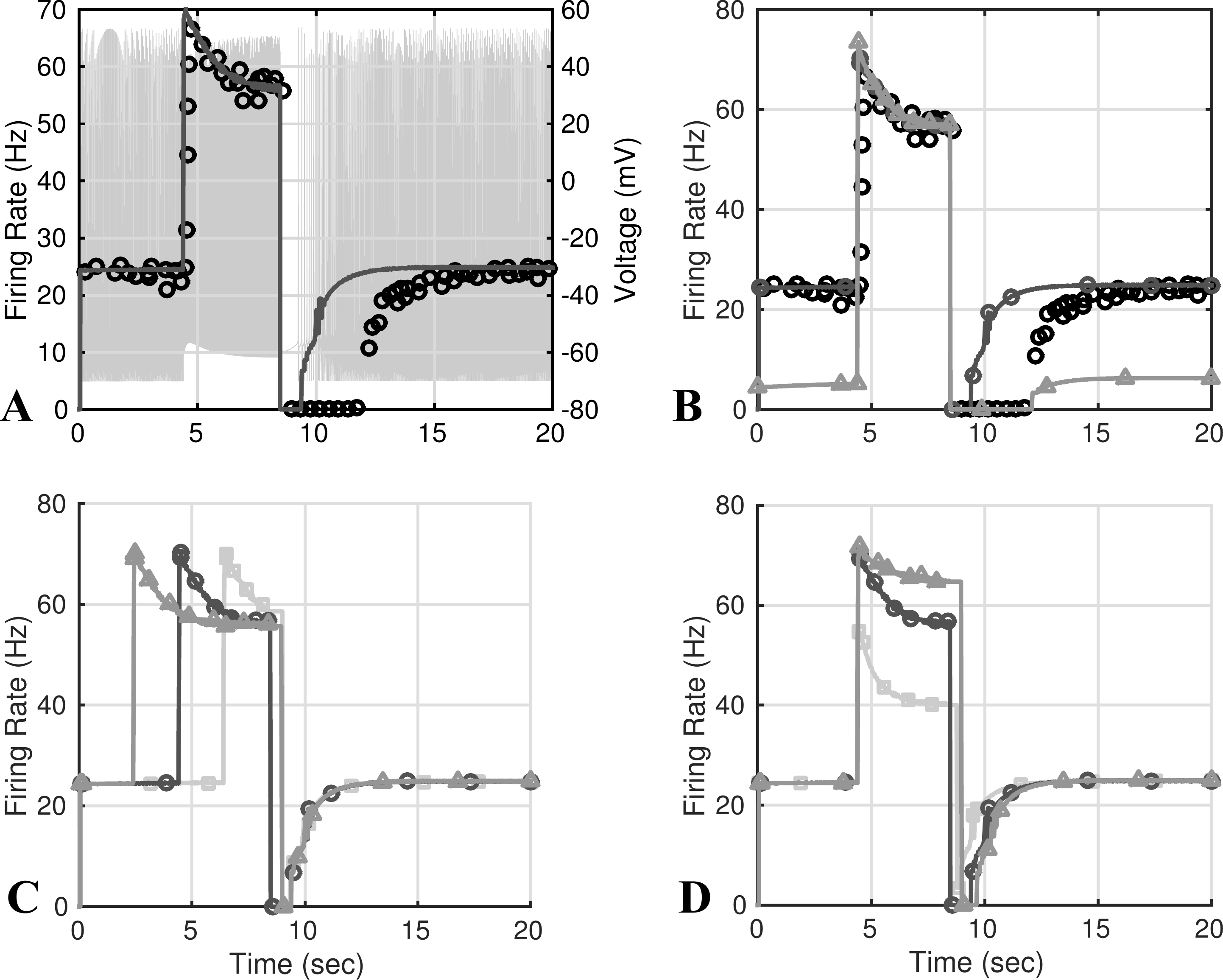}
\caption{ Using parameters estimated in response to the step stimulus used in \fref{fig:StepResults}, we were able to reproduce PED. We stimulated the model with a pressure pulse changing from 120 to 156 mmHg with a duration of 4.1 sec and compared the predicted firing rate to that of rat A-type neurons reported by \cite{saum_electrogenic_1976}. The computed and experimental firing rates are shown in \textbf{A}.  \textbf{B} shows results of varying \gm{}, $\epsilon_{\nicefrac{1}{2}}$, $S_{\nicefrac{1}{2}}$, and $\beta_2$  to match the duration of the PED. These parameter changes results in a lower the baseline firing rate (lighter curve with triangle markers).  \textbf{C} and \textbf{D} show results of increasing the length (2.1 (squares), 4.1 (circles), and 6.1 (triangles) sec) and amplitude (15 (squares), 20 (circles), and 50 (triangles) mmHg) of the pressure pulse.}
\label{fig:PulseResponse}
\end{figure*}

\begin{figure}
\centering
\includegraphics[width=.45\textwidth]{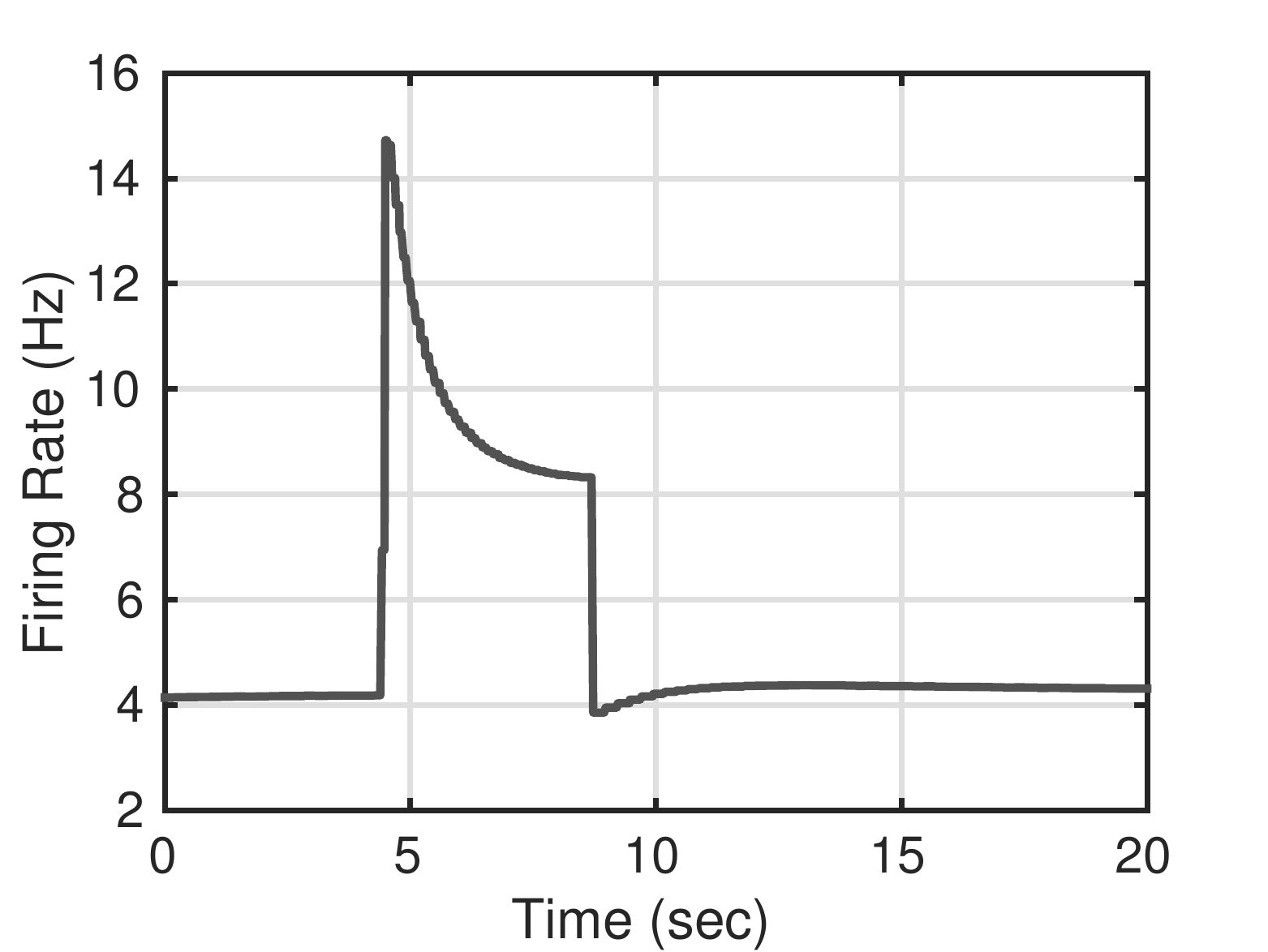}
\caption{Using parameters estimated from fitting the C-type ramp response in \fref{fig:RampResults}, we simulated the C-type neuron's response to a pressure pulse stimulus of the same shape used in \fref{fig:PulseResponse}. The results show a very peaked overshoot, with a complete lack of PED following the return to baseline pressure.}
\label{fig:CTypePulse}
\end{figure}

\paragraph{Pulse stimulus:}
To investigate the ability to characterize PED, the model was stimulated by a pressure pulse~\eqref{eq:pulse} with $p_{b} = 120$ mmHg, $\Delta p = 36$ mmHg, $t_{\mathrm{up}}=4.5$ sec, and $t_{\mathrm{down}} = 8.6$ sec. These values were chosen to match the data from \cite{saum_electrogenic_1976}, but with pressure values shifted to match the step stimulus used in previous simulations. Parameters were estimated minimizing \eqref{eq:lsqObj} using the  Nelder-Mead algorithm in \emph{JSIM}. The resulting response is shown in \fref{fig:PulseResponse}A. While simulations were able to predict the baseline, overshoot, and adaptation, the duration of the PED was too short. PED can be extended to match the data (see \fref{fig:PulseResponse}B), but this is at the cost of lowering the baseline firing rate.  To compare the model's behavior to that recorded by \cite{landgren_excitation_1952}, we simulated PED with pressure step durations 2.1, 4.1 and 6.1 sec, and amplitudes of 15, 20 and 50 mmHg, see Figs. \ref{fig:PulseResponse}C and D, respectively. Finally, we used the same stimulus to simulate a C-type neuron's response to a pulse pressure stimulus, the resulting firing rates are shown in \fref{fig:CTypePulse}.  To our knowledge no recordings of C-type baroreceptors' firing rate response to a pulse stimulus have been made.

\begin{figure}
\centering
\includegraphics[width=.45\textwidth]{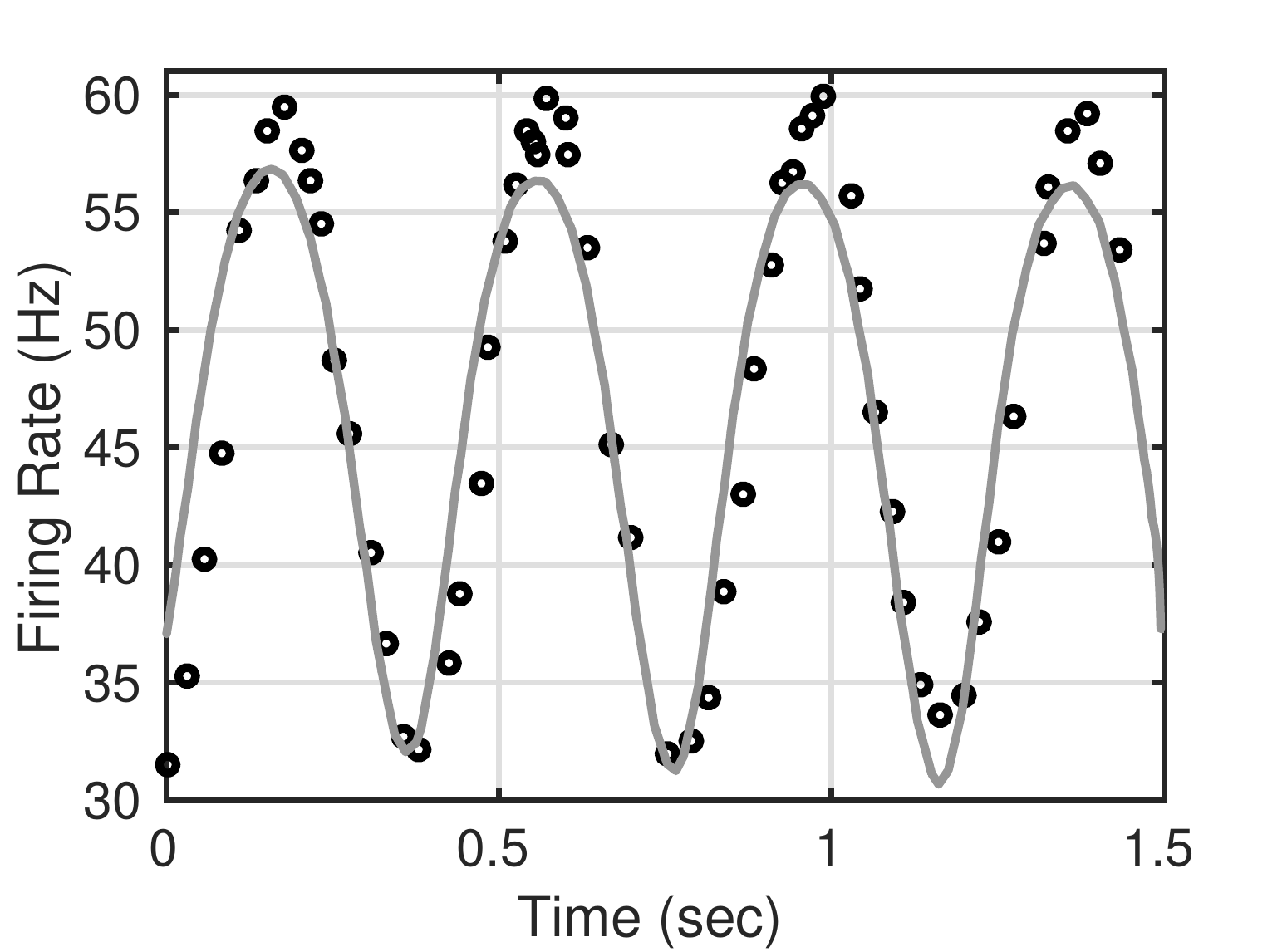}
\includegraphics[width=.45\textwidth]{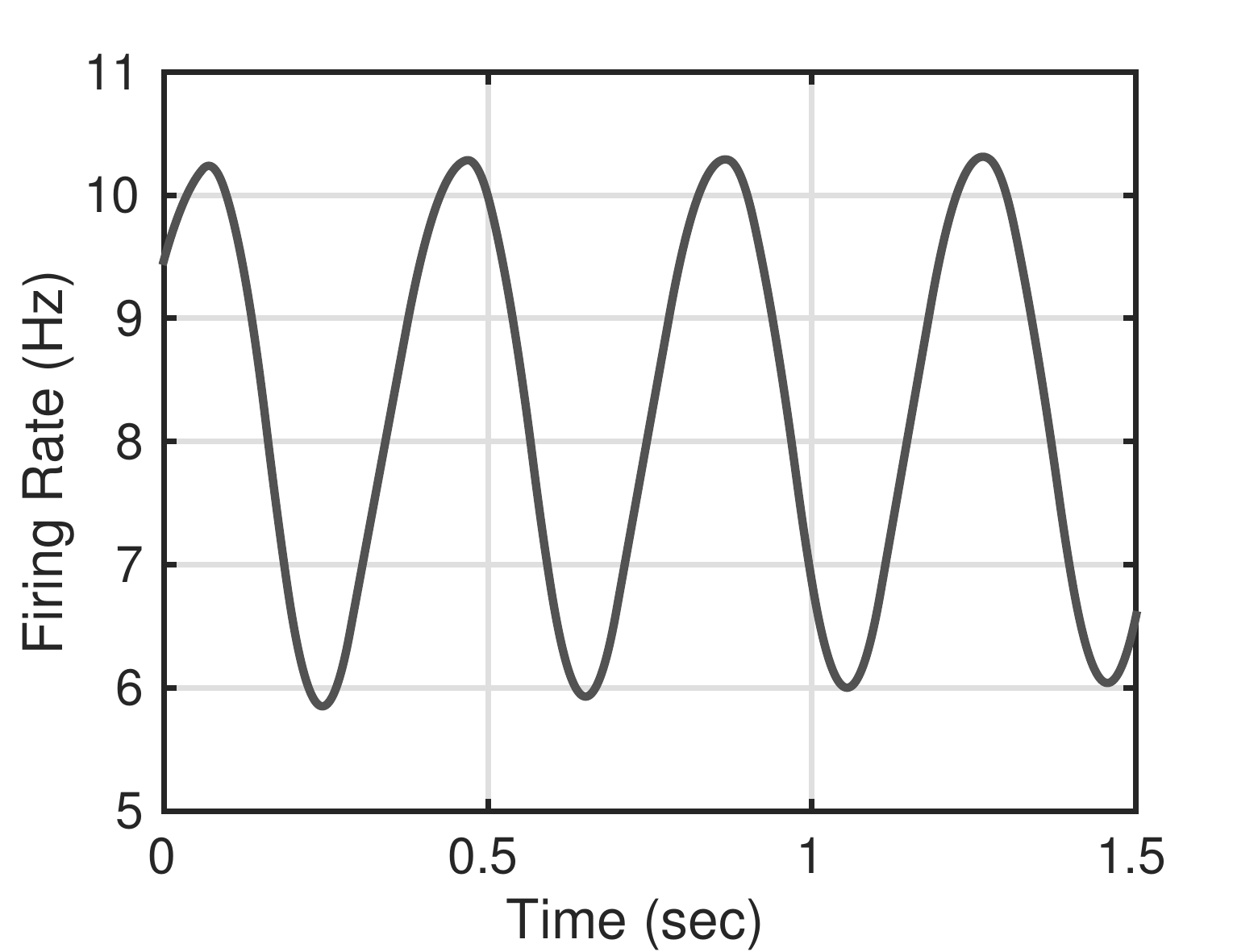}
\caption{Firing rate response to a sinusoidal input \eqref{eq:sineStim} with amplitude $p_A = 12.5$ mmHg, mean $p_b=140$ mmHg, frequency $\omega=2.5$ Hz, and  phase shift $\phi = -0.1$. The shape of the curve closely fits the experimental data reported by \cite{franz_small_1971} for A-type rabbit neurons, though the stimulus $p_b$ was adjusted to a higher level than that used to produce the experimental recordings. The resulting estimates of parameters are given in~\tref{tab:ParsResults}.}
\label{fig:SineResults}
\end{figure}

\paragraph{Sinusoidal stimulus:} To fit the firing rate response to sinusoidal data for A-type neurons recorded by \cite{franz_small_1971}, we used a stimulus with an amplitude  $p_A = 12.5$ mmHg, a mean pressure $p_b=140$ mmHg, a frequency $\omega=2.5$ Hz, and a phase shift $\phi = -0.1$.  For this simulation parameters were estimated using the Nelder-Mead algorithm in \emph{JSIM}.  This stimulus corresponds in shape and amplitude to that used to record the data but shifted in mean pressure in order to allow a common set of neuronal parameters to be used.  In addition to fitting the model parameters to A-type data, we also used the estimated parameters from the C-type ramp data (see \fref{fig:RampResults}) to simulate the firing rate response of a C-type neuron to the same sinusoidal stimulus. The resulting firing rates and parameters are shown in \fref{fig:SineResults}.

\section{Discussion}

The objective of this study was to develop a biophysical model of baroreceptor transduction of blood pressure that can reproduce differentiated responses of A- and C-type baroreceptor neurons.  
This extends existing baroreceptor models (\cite{alfrey_model_1997,bugenhagen_identifying_2010,mahdi_modeling_2013}), limited by using either \textit{ad hoc} models  (\cite{bugenhagen_identifying_2010,mahdi_modeling_2013}) or by focusing on prediction of A-type dynamics (\cite{alfrey_model_1997}). 

The model has three components corresponding to the  physical processes involved in the transduction of blood pressure: arterial wall deformation, nerve ending deformation and stimulation of mechanosensitive channels, and generation of action potentials by the neuronal membrane's ion-channel dynamics.  
The firing rate of the baroreceptor neurons  firing rate was calculated from interspike intervals of the membrane voltage. 
The ion channels were characterized according to results of previous studies of baroreceptors' electrophysiological characteristics. 
Using this model, we were able to, both quantitatively and qualitatively, reproduce known differences between A- and C-type signaling. 
This was done by adjusting the parameters of the mechanical coupling ($\alpha_1$, $\alpha_2$, $\beta_1$, $\beta_2$, $\epsilon_{\nicefrac{1}{2}}$, $S_{\nicefrac{1}{2}}$), variations of which could represent differences in the mechanosensitive propreties of different individual neurons due to variability of the anatomy of the nerve endings as well as differences between A- and C-type mechanosensitivity. 
Additionally, allowing $E_m$ to vary improved the fits shown in \fref{fig:PulseResponse} and \fref{fig:SineResults}, which may be interpreted as accounting for variations in the relative permeability of the mechanosensitive channel to certain ions or changes in the ionic concentrations. 
Changing the relative expressions of the fast sodium channel, the delayed rectifier potassium channel, the 4-AP sensitive potassium channels, the mechanosensitive channel and the sodium background current (\gNaF{}, \gKdr{}, \gKA{}, \gKD{}, \gm{}, and \gNaB{}) also contributed to reproducing the different firing patterns in the data. 
The differences in these values may be attributed to individual neuron variation, differences between A- and C-type neurons, and interspecies variability. 
For each data set, the estimated parameter values are given in \tref{tab:ParsResults}.

The parameters within the arterial wall deformation model were fit to data from \cite{feng_theoretical_2007} and subsequently used in all simulations. Since, to our knowledge, no data exist showing nerve ending deformation relative to the wall deformation, the nerve ending deformation model was calibrated as suggested in a previous modeling study (\cite{mahdi_modeling_2013}), which used an integrate and fire model to predict the firing rate. We further adjusted the activation parameters, $\epsilon_{\nicefrac{1}{2}}$ and $S_{\nicefrac{1}{2}}$ within the mechanosensitive ion channel and the maximal conductances, $g_{i}$, for each ion channel to reproduce available firing rate data extracted from published experiments (see Section~\sref{sec:experimental_data}). Other parameter values used in model of action potential generation were taken from the previous study by \cite{schild_-_1994}.

The firing patterns for A- and C-type neurons are typically distinguished by their unique response to a ramp pressure stimulus (\cite{seagard_firing_1990,coleridge_characteristics_1987,gilmore_comparison_1984}). A-type neurons have a distinct threshold pressure below which they do not fire, and they have a relatively large minimum firing rate (i.e. $>$ 10 Hz). C-type neurons, on the other hand, may be active over ``all'' pressure ranges, but have a pressure threshold at which their firing rate begins to increase in response to an increase in pressure. Their firing rage is lower than A-type neurons (between 2 and 20 Hz). These differences were reproduced in \fref{fig:RampResults} by changing the relative expression of potassium currents and the strength of the mechanosensitive current. Our results agree with observations by \cite{schild_differential_2012} who reported different levels of expression of potassium currents.  However, no data are available quantifying differences in the mechanosensitivity between the two neuron types. Our results of fitting ramp data suggest that the C-type neurons have a lower level of current carried by the mechanosensitive pathway. This difference may be due to a difference of ion channel expression between the two neuron types, such that C-type neurons have a significantly lower density of mechanonsensitive channels in the terminal endings than the myelinated A-type neurons.  Further since the C-type nerve endings are generally smaller and thus have lower surface area as reported by Krauhs (1979), the total conductance for a given channel density would also be lower. Fitting the various published experimental data by adjusting only maximal whole cell conductances and mechanical coupling parameters supports the explanation of the differences in these parameters as a reflection of neuroanatomical differences between A- and C-type terminal endings. 

The difference in maximal conductance could also be attributed to differences in the channel proteins expressed in A- and C-type vagal afferent neurons and baroreceptor terminal endings (presumably mechanosensitive). The possibility of differential protein expression in A- and C-type neurons is supported by the observations of~\cite{doan_chanel_dist_2004}, who found differential expression of HCN channels between A- and C-type neurons. It should be noted that no specific studies have shown that HCN channels in baroreceptor nerve endings are mechanosensitive, thus we speculate a similar difference in expression of the proteins composing the mechanosensitive channels. \cite{doan_chanel_dist_2004} also found that blocking HCN reduced the current threshold of the nerve endings, though they were not able to prove that the observed response was independent of muscle responses to the solution used to block HCN.  This finding in conjunction with the observation that C-type nerve endings have a lower HCN suggests that our observation of lower mechanosensitive conductance in C-type nerve endings may fit with the lower required current to activate nerve endings with lower HCN expression. Further, the higher expression of an HCN current with a faster time course in A-type nerve fibers may be a part of the explanation of how A-type nerve fibers achieve much higher firing rates than C-type nerve fibers.

To our knowledge, the only reported difference between A- and C-type baroreceptors in the response to a step stimulus are that A-type neurons typically have a higher firing rate of A-type, while C-type neurons display greater irregularity~\cite{brown_baroreceptor_1978}.  To match the dynamics displayed by A-type neurons, the ion channel conductance was increased, especially the sodium conductance (see  \fref{fig:StepResults}). Due to a lack of data for the step response for C-type baroreceptors, we were unable to test the model's ability to reproduce C-type data. However, we simulated a C-type response to step pressure stimulus using parameters estimated for the C-type ramp (see \fref{fig:rampStepResp}). The response obtained agrees with the qualitative features of overshoot and adaptation reported by \cite{brown_baroreceptor_1978}.

To better understand the differences associated with the individual model components, we considered how changes in individual model parameters effected the model output and found that increases in potassium conductances (\gKdr{}, \gKA{}, and \gKD{}) generally decreased the basal firing rate, with a smaller effect on maximal firing rates. 
These currents may be a key determinant of the threshold pressure and necessary for converting a constantly firing C-type neuron to one that fires only above a certain pressure stimulus. 
Raising \gm{} had the greatest effect on changing maximal firing rate, and sensitivity, though it had little effect on basal firing rate. 
Changing $S_{\nicefrac{1}{2}}$ and $\epsilon_{h}$  change the slope and threshold of the static pressure firing rate response observed in response to a ramp stimulus, but do not have an effect on its baseline or saturation. 
Increasing the half activation strain, $\epsilon_{\nicefrac{1}{2}}$ had a significant effect on the amount of adaptation observed in response to a pressure step stimulus. 
The parameters of the mechanical coupling have some effect on the baseline firing rate, though their primary effects are in the relation to the shape of the adaption curve in a step response. 
Increases in $\beta_{1}$ or $\beta_{2}$ result in a higher basal firing rate and a decreased scale of adaptation. 
Increases of $\alpha_{1}$ or $\alpha_{2}$ on the other hand have opposite effects.

These differences suggest that in C-type neurons the lower potassium conductance allows for sustained basal firing, whereas the higher mechanosensitive conductance is primarily responsible for the greater firing rate observed in A-type neurons. The ten fold difference between A-type and C-type \gm{} could be attributed to the smaller size of C-type axons and nerve endings, thus likely having a lower maximal conductance. The mechanical parameters of the C-type neuron on the other hand correspond to a greater scale of adaptation, which is particularly evident in \fref{fig:CTypePulse}. The interactions of these parameters are also important. A large change in the potassium conductance within C-type neurons can make them stop firing, and a large increase in $\gm$  results in cessation of firing above a certain level of stimulation.

Next, we tested the models' ability to predict and modulate PED in response to changes in the duration and amplitude of the step response \fref{fig:PulseResponse}, features that, to our knowledge, has not been discussed in previous modeling studies. While the model was able to elicit PED we were not able to match all features reported in the experiment. In particular we were not able to match the duration of the PED for the given baseline and pressure step amplitude. The model could predict a longer PED but at the cost of lowering the baseline firing rate (see \fref{fig:PulseResponse}B). As suggested by \cite{landgren_excitation_1952} we also tested the model's ability to predict a longer PED in response to a longer or higher step. While the predicted PED was longer, results were insignificant as shown in Figs. \ref{fig:PulseResponse}C and D. This could be due to our deterministic approach to simulating neural firing or to computations in a single neuron. It is well known that introducing small amounts of noise in neural models and to include interactions among neurons can change the dynamic behavior (\cite{McDonnell_2011, Solanka_2015}). However, it could also be due to mis-specification of the model predicting deformation of the neuron ending, or a result of excluded channels. 

Finally, we increased fast sodium conductance and decreased the background sodium current to reproduce the sinusoidal firing rate response data by \cite{franz_small_1971}. We simulated the C-type response to sinusoidal pressure  was simulated using the parameters estimated for the ramp fit. Results of this modulation were in qualitative agreement with features reported by \cite{brown_baroreceptor_1978}.

These quantitative and qualitative results indicate that a biophysical approach may indeed account for observed differences between A- and C-type neurons, primarily at the neuronal level due to electrophysiological differences. To our knowledge this is the first study that has reproduced both A- and C-type afferent firing dynamics with a single model.  The predictions of  the C-type response to step and pulse stimuli suggested that C-type fibers may not exhibit PED, a possibility that to our knowledge has not been discussed in previous studies. Furthermore, we predicted  the C-type  response to step and pulse stimuli, which provide an  explanation for the characteristics of these responses, as well as suggesting that C-type fibers do not exhibit PED, a characteristic which has not been discussed in previous studies. 

In addition to these differences, the model successfully reproduces the step response characteristics of the firing rate response of baroreceptors. These results are expected as we have developed a model that attempts to incorporate the characteristics of the underlying systems generating the firing rate response: the arterial wall mechanics and action potential generation in the afferent fiber endings. We observe that the variations in $\alpha_1$, $\alpha_2$, $\beta_1$, and $\beta_2$ seem to be greater within the group of A-type estimates as compared to those between A- and C-type ramp estimates (see \tref{tab:ParsResults}), thus we hypothesize that the mechanical coupling of the fiber endings is not significantly differentiated between A- and C-type nerve endings.
In fact these variations are quite large (2-3 fold) within the same nerve ending type. We hypothesize that this variability may be due to the convoluted and variable anatomical structure of the nerve endings as reported by \cite{krauhs_structure_1979}.
In addition, this variability could suggest that a single set of coupling parameters may not accurately describe the coupling of sensory endings in general. 

The data available span a large range of firing rate responses indicating significant variability within individual baroreceptor types. Our results show that modeling can provide a way to investigate differences between baroreceptor types, although limited to considering autoactive C-type fibers. Although this model of autoactive C-type fibers may not explain the irregular firing observed in non-autoactive C-type fibers, it may be useful to to explain differences between A- and Ah- type neurons studied by \cite{li_electrophysiological_2008}. The latter is of particular importance as this neuron-type may be essential to better understand pathophysiology associated in patients experiencing orthostatic intolerance.

One difficulty in this study is the lack of a data set containing recordings from a single nerve fiber's firing patterns in response to each of the stimuli considered in this study. Such a recording would ideally allow for a more consistent set of parameters to be used when reproducing each of the known firing patterns. The majority of firing rate recordings in response to a controlled pressure stimulus seem to have been conducted on A-type nerve fibers, making it difficult to evaluate the model's ability to capture the key features of C-type firing patterns. In addition, the lack of measurements of the coupling of the mechanosensitive currents to arterial wall deformation impedes better calibration and evaluation of the model's representation of these components.

Finally, our study included only the largest inward and outward currents reported in previous electrophysiology and modeling studies in order to simplify the analysis.  Other ion channels may play a significant role in long-term adaptation and pharmacological sensitivities of the baroreceptor firing patterns. For example, \cite{Gallego_AvsC_1978} demonstrated that the Nav1.8 (TTX-R) channel enabled vagal afferents to conduct and produce action potentials in the presence of TTX, although not without pronounced changes in electrical threshold and action potential shape. Likewise they showed that 
BK-type calcium activated potassium  (KCa1.1) is markedly expressed in unmyelinated vagal afferents and has been shown to be a robust modifier of neural discharge. Further studies should investigate the impact of adding more channels to study how these impact observed dynamics. In addition, various studies have identified specific ion channels using pharmacological, immunohistochemical, and genetic techniques \cite[Table 1]{schild_differential_2012}. These are examples of  voltage-gated and calcium-activated ion channels, which carry potassium, so\-di\-um, and calcium currents are found in baroreceptor neurons though their role in determining the firing rate is not described in detail.  It is possible that some of these channels contribute to dynamic fluctuations in calcium concentrations and thus contribute to frequen\-cy adaptation characteristics observed by \cite{brown_baroreceptor_1978}.

In summary, this study demonstrates the feasibility of a biophysical approach to map the differentiation between A- and C-type baroreceptor firing patterns using a common mathematical model based on the underlying physiology of the transduction process. Further work is needed to develop a comprehensive biophysical representation of the origin of the various baroreceptor firing characteristics, allowing for quantitative attribution of emergent firing rate features to particular variations in model parameters. Such an approach would provide a biophysical context for evaluating afferent baroreflex dysfunction, as this type of model would allow investigation of how physiological abnormalities may give rise to questions in the transduction of blood pressure. This model could be used to understand how selective inhibition of A- or C-type might occur and give rise to baroreflex dysfunction, which could correspond to unique etiologies of disorders such as orthostatic intolerance. 

\clearpage 

\begin{table*}[!t]
  \caption{Parameters used to fit the various stimuli shown in \fref{fig:RampResults} and \fref{fig:StepResults}. Neural parameters not estimated are reported in~\tref{tab:ParsDesc}. The root mean squared error (RMSE) of the model prediction compared to the data is also reported.}
\label{tab:ParsResults}
\centering {\footnotesize 
\begin{tabu}{l | c c c c c}
Simulation                  & A-step          & A-pulse         & A-sine          & A-ramp          & C-ramp          \\ \hline
$ \alpha_{1} $              & 5.794\sciE{-4}  & 5.794\sciE{-4}  & 5.804\sciE{-4}  & 1.1550\sciE{-4} & 1.1712\sciE{-4} \\
$\alpha_{2}$                & 4.00e\sciE{-4}  & 4.000e\sciE{-4} & 3.976\sciE{-4}  & 3.2473\sciE{-4} & 5.2057\sciE{-4} \\
$\beta_{1} $                & 5.2012\sciE{-4} & 5.2012\sciE{-4} & 5.255\sciE{-4}  & 3.4971\sciE{-4} & 2.0641\sciE{-4} \\
$ \beta_{2}$                & 2.00\sciE{-3}   & 2.000\sciE{-3}  & 2.000\sciE{-3}  & 9.8326\sciE{-4} & 2.500\sciE{-3}  \\
$\epsilon_{\nicefrac{1}{2}}$            & 1.85\sciE{-1}   & 2.10\sciE{-1}   & 1.85\sciE{-1}   & 2.72\sciE{-1}   & 3.048\sciE{-1}  \\
$S_{\nicefrac{1}{2}}$                   & 2.13\sciE{-2}   & 2.13\sciE{-2}   & 2.88\sciE{-2}   & 2.95\sciE{-2}   & 2.46\sciE{-2}   \\
\gNaF                       & 8.9230          & 8.9230          & 10.0197         & 2.0500          & 2.0500          \\
\gKdr                       & 9.90 \sciE{-3}  & 9.90 \sciE{-3}  & 9.90 \sciE{-3}  & 9.90 \sciE{-3}  & 5.50 \sciE{-3}  \\
\gKA                        & 1.68\sciE{-1}   & 1.68\sciE{-1}   & 1.68\sciE{-1}   & 6.30\sciE{-2}   & 3.50 \sciE{-2}  \\
\gKD                        & 1.80  \sciE{-2} & 1.80  \sciE{-2} & 1.80  \sciE{-2} & 1.80  \sciE{-2} & 1.80  \sciE{-2} \\
\gm                         & 2.3 \sciE{-3}   & 3.0 \sciE{-3}   & 2.3 \sciE{-3}   & 1.2 \sciE{-3}   & 1.0 \sciE{-4}   \\
$C_{m}$                     & 3.25\sciE{-2}   & 3.25\sciE{-2}   & 3.25\sciE{-2}   & 3.25\sciE{-2}   & 3.25\sciE{-2}   \\
$E_{\mathrm{m}}$            & 0.0             & 5.0             & 5.05            & 0.0             & 0.0             \\
\gNaB                       & 3.25\sciE{-4}   & 4.95\sciE{-4}   & 3.253\sciE{-4}  & 3.25\sciE{-4}   & 3.25\sciE{-4}   \\ \hline
RMSE                        & 4.611           & 14.05           & 2.83            & 3.565           & 1.062           \\ 
$R^2$                       & 0.8794          & 0.5793 & 0.8794 & 0.6904 & 0.9394 \\ \hline
\end{tabu}}
\end{table*}

\begin{table*}[!t]
\centering
\caption{ Description of the state variables and auxillary quantities used in this paper.} \label{tab:VarsDesc}
{\footnotesize
\begin{tabular}{c l c}
\textbf{Variable}                                                                     & \hspace{0.5cm} \textbf{Definition}           & \textbf{Units}     \\ \hline
$p$                                                                                   & aortic blood pressure                        & mmHg               \\
$\epsilon_w$                                                                          & aortic wall strain                           & unitless           \\
$\epsilon_1$                                                                          & nerve ending coupling strain 1               & unitless           \\
$\epsilon_2$                                                                          & nerve ending coupling strain 2               & unitless           \\
$\epsilon_{ne}$                                                                       & nerve ending strain                          & unitless           \\
$V$                                                                                   & membrane voltage                             & mV                 \\
$m$                                                                                   & Nav1.7 Activation                            & unitless           \\
$h$                                                                                   & Nav1.7 Inactivation                          & unitless           \\
$j$                                                                                   & Nav1.7 Reactivation                          & unitless           \\
$n$                                                                                   & Delayed Rectifier activation                 & unitless           \\
$p$                                                                                   & K A Activation                               & unitless           \\
$q$                                                                                   & K A Inactivation                             & unitless           \\
$x$                                                                                   & K D Activation                               & unitless           \\
$y$                                                                                   & K D Inactivation                             & unitless           \\
$f$                                                                                   & firing rate                                  & Hz                 \\
                                                                                      &                                              &                    \\
                                                                                      &                                              &                    \\
\textbf{Auxillary Quantites}                                                          & \hspace{0.5cm} \textbf{Definition}           & \textbf{Units}     \\ \hline
\INaF                                                                                 & Nav1.7 Current                               & nA                 \\
\IKdr                                                                                 & K-DR current                                 & nA                 \\
\IKA                                                                                  & K-A current                                  & nA                 \\
\IKD                                                                                  & K-D current                                  & nA                 \\
\IM                                                                                   & MSC current                                  & nA                 \\
\(p_{o} \)                                                                            & open probability of MSC                      & unitless           \\
\INaB                                                                                 & Sodium background leakage                    & nA                 \\
\ICaB                                                                                 & Calcium background leakage                   & nA                 \\
\INaK                                                                                 & Sodium-potassium exchanger current           & nA                 \\
\INaCa                                                                                & Sodium-calcium exchanger current             & nA                 \\
\ICaP                                                                                 & Sodium-potassium pump current                & nA                 \\
\end{tabular}}
\end{table*}

\begin{table*}[!t]
\centering
\caption{Description of the parameters used in this paper and their nominal values.}
\label{tab:ParsDesc}
{\footnotesize
\begin{tabular}{c l c c c}
\textbf{Parameter}                                                                    & \hspace{0.5cm} \textbf{Definition}           & \textbf{Value}     &  \textbf{Units}                                     &  \textbf{Reference}                    \\ \hline
$A_0$                & unstressed aortic area                       & $4.01$             & $\mathrm{mm}^2$                                    &  \cite{valdez-jasso_viscoelastic_2009} \\
$A_m$                & maximal aortic area                          & $15.708$           & $\mathrm{mm}^2$                                    &  \cite{valdez-jasso_viscoelastic_2009} \\
$\alpha$             & saturation pressure                          & $145$              & mmHg                                               &  \cite{valdez-jasso_viscoelastic_2009} \\
$\kappa_w$           & steepness const                              & $5$                & unitless                                           &  \cite{valdez-jasso_viscoelastic_2009} \\
$R_{A}$              & maximal to minimal area ratio                &                    & unitless                                           &                                        \\
$E_0$                & elastic nerve const                          & $1$                & mmHg                                               &  \cite{bugenhagen_identifying_2010}    \\
$E_1$                & elastic nerve const                          & $1$                & mmHg                                               &  \cite{bugenhagen_identifying_2010}    \\
$E_2$                & elastic nerve const                          & $5$                & mmHg                                               &  \cite{bugenhagen_identifying_2010}    \\
$\eta_1$             & viscous nerve coupling const                 & $2$                & mmHg $\cdot$ s                                     &  \cite{bugenhagen_identifying_2010}    \\
$\eta_2$             & viscous nerve coupling const                 & $2.5$              & mmHg $\cdot$ s                                     &  \cite{bugenhagen_identifying_2010}    \\
$\alpha_1$           & nerve ending const                           & $E_0/\eta_1$       & $\mathrm{s}^{-1}$                                  &  \cite{mahdi_qualitative_2012}         \\
$\alpha_2$           & nerve ending const                           & $E_0/\eta_2$       & $\mathrm{s}^{-1}$                                  &  \cite{mahdi_qualitative_2012}         \\
$\beta_1$            & nerve ending relaxation rate                 & $E_1/\eta_1$       & $\mathrm{s}^{-1}$                                  &  \cite{mahdi_qualitative_2012}         \\
$\beta_2$            & nerve ending relaxation rate                 & $E_2/\eta_2$       & $\mathrm{s}^{-1}$                                  &  \cite{mahdi_qualitative_2012}         \\
$s_1$                & firing const                                 & 480                & $\mathrm{s}^{-1}$                                  &  \cite{mahdi_modeling_2013}            \\
$s_2$                & firing const                                 & 100                & $\mathrm{s}^{-1}$                                  &  \cite{mahdi_modeling_2013}            \\
$C_m$                & membrane capacitance                         & 32.5               & pF                                                 &  \cite{schild_-_1994}                  \\
$E_{Na}$             & sodium reversal potential                    & 72.8               & mV                                                 &  \cite{schild_-_1994}                  \\
$E_{K}$              & Potassium reversal potential                 & -83.9              & mV                                                 &  \cite{schild_-_1994}                  \\
$E_{Ca}$             & Calcium reversal potential                   & 126.7              & mV                                                 &  \cite{schild_-_1994}                  \\
$g_{Nav1.7}$         & maximal Nav1.7 conductance                   & 2.05               & \(\mu \mathrm{S}\)                                 &  \cite{schild_-_1994}                  \\
$g_{K,DR}$           & maximal delayed rectifier conductance        & 0.0055             & \(\mu \mathrm{S}\)                                 &  \cite{schild_-_1994}                  \\
$g_{K,A}$            & maximal transient 4AP sensitive conductance  & 0.035              & \(\mu \mathrm{S}\)                                 &  \cite{schild_-_1994}                  \\
$g_{K,D}$            & maximal persistent 4AP sensitive conductance & 0.0100             & \(\mu \mathrm{S}\)                                 &  \cite{schild_-_1994}                  \\
$g_{Na,B}$           & Background sodium conductance                & 3.25e-4            & \(\mu \mathrm{S}\)                                 &  \cite{schild_-_1994}                  \\
$g_{Ca,B}$           & Background calcium conductance               & 8.25e-5            & \(\mu \mathrm{S}\)                                 &  \cite{schild_-_1994}                  \\
$g_{m}$              & mechanosensitive channel conductance         & 1.00e-4            & \(\mu \mathrm{S}\)                                 &  \cite{alfrey_model_1997}              \\
$\epsilon_{\nicefrac{1}{2}}$     & half activation nerve strain                 & 0.3048             & unitless                                           &  \cite{alfrey_model_1997}              \\
$S_{\nicefrac{1}{2}}$            & reciprocal slope for \(p_{o}\)               & 0.0246             & unitless                                           &  \cite{alfrey_model_1997}              \\
\(\bar{I}_{Na,K}\)   & Maximal sodium-potassium exchanger current   & 0.275              & nA                                                 &  \cite{schild_-_1994}                  \\
\(\bar{I}_{Ca,P}\)   & Maximal calcium pump current                 & 0.0243             & nA                                                 &  \cite{schild_-_1994}                  \\
\(\Ion{Na}{+}_i\)    & Intracellular sodium concentration           & 8.90               & mM                                                 &  \cite{schild_-_1994}                  \\
\(\Ion{Na}{+}_o\)    & Extracellular sodium concentration           & 154                & mM                                                 &  \cite{schild_-_1994}                  \\
\(\Ion{K}{+}_i\)     & Intracellular potassium concentration        & 145                & mM                                                 &  \cite{schild_-_1994}                  \\
\(\Ion{K}{+}_o\)     & Extracellular potassium concentration        & 5.40               & mM                                                 &  \cite{schild_-_1994}                  \\
\(\Ion{Ca}{+}_i\)    & Intracellular calcium concentration          & 9.70e-05           & mM                                                 &  \cite{schild_-_1994}                  \\
\(\Ion{Ca}{+}_o\)    & Extracellular calcium concentration          & 2.0                & mM                                                 &  \cite{schild_-_1994}                  \\
$D_{\mathrm{NaCa}}$  &                                              & 0.0036             & mM$^{-4}$                                          &  \cite{schild_-_1994}                  \\
$K_{\mathrm{M,K}}$   &                                              & 0.6210             & mM                                                 &  \cite{schild_-_1994}                  \\
$K_{\mathrm{M,Na}}$  &                                              & 5.46               & mM                                                 &  \cite{schild_-_1994}                  \\
$K_{\mathrm{M,CaP}}$ &                                              & 0.00005            & mM                                                 &  \cite{schild_-_1994}                  \\
$K_{\mathrm{NaCa}}$  &                                              & $3.6\times10^{-5}$ & nA $\cdot$ mM$^{-4}$                               &  \cite{schild_-_1994}                  \\
$\gamma$             & exchange ratio                               & 0.5                &                                                    &  \cite{schild_-_1994}                  \\
$r$                  & exchange ratio                               & 3                  &                                                    &  \cite{schild_-_1994}                  \\
$R$                  & Ideal gas const                              & 8314               & $\sfrac{\mathrm{J}}{\mathrm{mol}\cdot \mathrm{K}}$ \\
$F$                  & Faraday's const                              & 96500              & $\sfrac{\mathrm{C}}{\mathrm{mol}}$                 \\ 
\end{tabular}}
\end{table*}

\clearpage

\section{Appendix}\label{sec:appendix}

\subsection{Model equations}
\subsubsection{Arterial wall deformation}
Nonlinear (sigmoidal) relation between vessel area $A$ and blood pressure $p$:
\begin{equation*}
     A(p)=(A_m-A_0)\frac{p^k}{\alpha_w^k+p^k}+ A_0.
\end{equation*}
Vessel strain $\epsilon_w$:
\begin{equation*}
    \epsilon_{w} = \frac{r - r_{0}}{r}=1-\sqrt{\frac{(\alpha_w^k+p^k)}{\alpha_w^k+R_{A} p^k}}.
\end{equation*}
where \( R_{A} = \sfrac{A_{m}}{A_{0}} \).

\subsubsection{Nerve ending deformation}
Neuron ending deformation $\epsilon_{ne}$ predicted using the two-element Voigt body model in 
\fref{fig:V2Coupling}:
\begin{eqnarray*}
    \frac{d\epsilon_1}{dt}&=&-(\alpha_1\!+\!\alpha_2\!+\!\beta_1)\epsilon_1 \!+\! (\beta_1\!-\!\beta_2)\epsilon_2\! \\
    && +\! (\alpha_1\!+\!\alpha_2)\epsilon_{w},\\
    \frac{d\epsilon_2}{dt}&=&-\alpha_2\epsilon_1 - \beta_2\epsilon_2  + \alpha_{2} \epsilon_{w},
\end{eqnarray*}
where
\begin{eqnarray*}    
   \epsilon_{ne} &=& \epsilon_w -\epsilon_1.
\end{eqnarray*}

\subsubsection{Mechanosensitive ionic current}
\begin{equation*}
     p_{o}(\epsilon_{ne}) = \left\{ 1 + \exp\left(\frac{\epsilon_{\nicefrac{1}{2}} - \epsilon_{ne}}{S_{\nicefrac{1}{2}}}\right)\right\}^{-1}.
\end{equation*}

\begin{equation*}
      \IM{} = p_o(\epsilon_{ne}) \gm{} (V-E_{\mathrm{msc}}).
\end{equation*}

\subsubsection{Conductance based model of afferent action potential generation}
The reversal potentials of the ions are calculated using the Nernst equation
\begin{equation*}
E_{X} = \frac{RT}{F} \ln{ \left( \frac{ [\text{X}^{+}]_\mathrm{out}}{ [\text{X}^{+}]_\mathrm{in}} \right)}.
\end{equation*}
The membrane voltage is governed by
\begin{equation*}
  \frac{dV}{dt} = -\sfrac{\sum I_{i}}{C_m}.
\end{equation*}

\noindent Fast sodium current:
\begin{align*}
\INaF{} = & \gNaF m^3 h j (V-E_{Na}), \\
\frac{dm_f}{dt} =& \ \frac{m_{\infty} - m}{\tau_{m}}, \\
\frac{dh_f}{dt}  =& \ \frac{h_{\infty} - h}{\tau_{h}}, \\
\frac{dj_f}{dt}   =& \ \frac{j_{\infty} - j}{\tau_{j}},
\end{align*}
and 
\begin{align*}
m_{\infty} (V) =& \frac{1}{1+\exp\left\{-(V+41.35)/4.75\right\}},\\
h_{\infty} (V)  =& \frac{1}{1+\exp\left\{ (V+62.00)/4.50\right\}},\\
j_{\infty} (V)   =& \frac{1}{1+\exp\left\{ (V+40.00)/1.50\right\}},\\
\tau_{m}(V)    =& 0.75 \exp\left\{-(0.0635)^2 (V+40.35)^2\right\}, \\
& + 0.12\\
\tau_{h}(V)     =& 6.50 \exp\left\{-(0.0295)^2 (V+75.00)^2\right\}, \\
& + 0.55\\
\tau_{j}(V)      =& \frac{25}{1+\exp\left\{ (V-20.00)/4.50\right\}} + 0.01.
\end{align*}

\noindent Delayed rectifier potassium current:
\begin{align*}
\IKdr{} =& \gKdr{} n (V-E_{\mathrm{K}}),\\
\frac{dn}{dt}     =& \frac{n_{\infty} - n}{\tau_{n}},\\
n_{\infty} (V)    =& \frac{1}{1+\exp\left\{(V+14.62)/18.38\right\}},\\
\tau_n(V)         =& \frac{1}{\alpha_n +\beta_n} + 1.0, \\
\alpha_n          =& \frac{0.001265 (V+14.273)}{1 - \exp \left\{-(V+14.273)/10\right\}},\\
\beta_n            =& 0.0125 \exp\left(\frac{-(V+55)}{2.5}\right).
\end{align*}

\noindent 4-AP sensitive potassium currents:
\begin{align*}
\IKA{} =& \gKA{} p^3 q(V-E_{\mathrm{K}}),\\
\frac{dp}{dt}     =& \frac{p_{\infty} - p}{\tau_{p}},\\
\frac{dq}{dt}     =& \frac{q_{\infty} - q}{\tau_{q}},\\
p_{\infty}             =& \left\{1+\exp(-(V+28.0)/28.0)\right\}^{-1}, \\
q_{\infty}             =& \left\{1+\exp((V+58.00)/7.0)\right\}^{-1} ,    \\
\tau_{p}               =& 5.0\exp\left\{-(0.022)^2 (V+65.0)^2  \right\} + 2.5,  \\
\tau_{q}               =& 100.0\exp\left\{-(0.035)^2 (V+30.00)^2  \right\}, \\
& + 10.5, 
\end{align*}
and
\begin{align*}
\IKD{}    =& \gKD{} x^3 y (V - E_{\mathrm{K}}), \\
\frac{dx}{dt}     =& \frac{x_{\infty} - x}{\tau_{x}},\\
\frac{dy}{dt}     =& \frac{y_{\infty} - y}{\tau_{y}},\\
x_{\infty}  =& \left\{1+\exp(-(V+39.59)/14.68)\right\}^{-1},  \\
y_{\infty}  =& \left\{1+\exp((V+48.00)/7.0)\right\}^{-1},   \\
\tau_{x}    =& 5.0\exp(-(0.022)^2 (V+65.0)^2) + 2.5, \\
\tau_{y}    =& 7500.0. 
\end{align*}

\noindent Leakage currents:
\begin{align*}
\INaB{} =& \gNaB{} (V-E_{Na}). \\
\ICaB{} =& \gCaB{} (V-E_{Ca}).
\end{align*}

\noindent Sodium-potassium exchange current:
\begin{equation*}
\INaK{} = \bar{I}_{NaK} \left( \frac{ \Ion{Na}{+}_{i} }{\Ion{Na}{+}_{i} + K_{M,Na}}\right)^{3} \left( \frac{ \Ion{K}{+}_{o} }{\Ion{K}{+}_{o} + K_{M,K}}\right)^{2}.
\end{equation*}

\noindent Sodium-calcium exchange current:
\begin{align*}
&S             = 1 + D_{NaCa}\left(\Ion{Ca}{+}_{i}\Ion{Na}{+}_{o}^3 +\Ion{Ca}{+}_o \Ion{Na}{+}_{i}^3\right),\\
&DF_{in}   = \Ion{Na}{+}_{o}^3 \Ion{Ca}{+}_o \exp(\gamma V F/ R T),\\
&DF_{out} = \Ion{Na}{+}_o^3 \Ion{Ca}{+}_i \exp\{(\gamma-1) V F/ R T \},\\
&\INaCa{} = K_{\mathrm{Na,Ca}} (DF_{\mathrm{in}}-DF_{\mathrm{out}}) / S.
\end{align*}

\noindent Calcium pump:
\begin{equation*}
\ICaP = \overline{I}_{\mathrm{CaP}} \frac{\Ion{Ca}{+}_{i}}{\Ion{Ca}{+}_{i}+\overline{K}_{Ca}}.
\end{equation*}

\clearpage

\bibliographystyle{aps-nameyear}      
\bibliography{bibliog}                
\nocite{*}

\end{document}